\documentclass[structabstract]{aa}	
\usepackage{graphicx}
\usepackage{txfonts}
\usepackage{natbib}
\bibpunct{(}{)}{;}{a}{}{,}

\begin{document}
\title{Heavy coronal  ions in the heliosphere} 
\subtitle{II. Expected fluxes of energetic neutral He atoms from the heliosheath}

\author{S.~Grzedzielski 
\and  P.~Swaczyna 
\and  M.~Bzowski  
}

\institute{Space Research Centre, Polish Academy of Sciences, Bartycka 18A, 00-716 Warsaw, Poland \email{stangrze@cbk.waw.pl} \label{inst1}}

\date{Received [date] / Accepted [date]}

\abstract
{}
{A model of heliosheath density and energy spectra of $\alpha$-particles and He$^+$ ions carried by the solar wind is developed. Neutralization of heliosheath He ions, mainly by charge exchange (CX) with neutral interstellar H and He atoms, gives rise to $\sim$0.2~--~$\sim$100~keV fluxes of energetic neutral He atoms (He ENA). Such fluxes, if observed, would give information about plasmas in the heliosheath and heliospheric tail.}
{Helium ions crossing the termination shock (TS) constitute suprathermal (test) particles convected by (locally also diffusing through) hydrodynamically calculated background plasma flows (three versions of flows are employed). The He ions proceed from the TS towards heliopause (HP) and finally to the heliospheric tail (HT). Calculations of the evolution of $\alpha$- and He$^+$ particle densities and energy spectra include binary interactions with background plasma and interstellar atoms (radiative and dielectronic recombinations, single and double CX, stripping, photoionization and impact ionizations), adiabatic heating (cooling) resulting from flow compression (rarefaction), and Coulomb scattering on background plasma.}
{Neutralization of suprathermal He ions leads to the emergence of He ENA fluxes with energy spectra modified by the Compton-Getting effect at emission and ENA loss during flight to the Sun. Energy-integrated He ENA intensities are in the range $\sim$0.05~--~$\sim$50~cm$^{-2}$s$^{-1}$sr$^{-1}$ depending on spectra at the TS (assumed kappa-distributions), background plasma model, and look direction. The tail/apex intensity ratio varies between $\sim$1.8 and $\sim$800 depending on model assumptions. Energy spectra are broad with maxima in the $\sim$0.2~--~$\sim$3~keV range depending on the look direction and model.}
{Expected heliosheath He ENA fluxes may be measurable based on the capabilities of the IBEX spacecraft. Data could offer insight into the heliosheath  structure and improve understanding of the post-TS solar wind plasmas. HT direction and extent could be assessed.}


\maketitle

\section{Introduction}
\label{introduction}
Helium, as the second abundant ion species in the solar wind, should also be prominent in energetic neutral atom (ENA) fluxes from the heliosheath, resulting from the transcharge on neutral atom populations. Detection of these fluxes is contingent upon the energy of the He ENAs. It is known that  He ions at interplanetary shocks do not equilibrate their downstream thermal energies with protons. SWICS experiment data obtained on board Ulysses s/c for 15 shocks at distances between 2.7 and 5.1 AU from the Sun indicate that the mean ratio of $\alpha$-to-protons downstream thermal velocities is $1.3\pm0.3$ \citep{berdichevsky_etal:97a}. High thermal speeds of $\alpha$-particles (and pickup He$^+$) at shocks accompanied by enhanced high-energy power-law tails were also seen by Ulysses SWICS and HISCALE experiments  in a reverse quasi-perpendicular CIR shock and also at the inbound pass of the quasi-parallel Jovian bow shock \citep{gloeckler_etal:05a}.
Instances of non-thermalization of He ions at Earth bowshock were reported and discussed by \citet{liu_etal:07a}. 

All this suggests that the He ions in the heliosheath can be treated at injection, i.e. immediately upon crossing the TS, as thermally separate from the bulk of shocked solar wind plasma: the average post-shock energy should be close to the upstream energy in a solar frame, with tail distributions approximated by power laws in energy. The excess energy, compared to ideal gas dynamics, of the random motions of He ions in the heliosheath can probably persist for a long time. 
The Coulomb equilibration time for 1~keV/n $\alpha$-particles (i.e. bulk pre-TS He) with heliosheath thermal protons of number density $\sim$0.001~cm$^{-3}$ and temperature $\sim$70\,000~K \citep{richardson:11a} amounts to 5$\,\times\,10^{11}$~s, which is longer than the heliosheath residence time $\sim$10$^9$~s. Loss of energy through interaction with neutral H atoms in the heliosheath is also slow. For total stopping power for 1~keV/n $\alpha$-particles in H equal to 1120~MeVcm$^2$g$^{-1}$ \citep{berger_etal:05a} the characteristic time for $\alpha$-energy decay in neutral H density $0.2\ \mathrm{cm}^{-3}$ is $\sim$2$\,\times\,10^{11}$~s.
All this suggests that He ions in the heliosheath will constitute a relatively hot particle population, with typical energies $\sim$0.5--1~keV/n, so well above the thermal proton plasma measured by Voyager, and with nonthermal tails reaching several keV/n. Neutralization of He ions, mainly by charge exchange with neutral interstellar H and He, should therefore lead to the emergence of fluxes of relatively energetic He ENAs. The aim of the present paper is to estimate the magnitude of theses fluxes and to look for the conditions for detection.

Our approach is in part similar to the one used by \citet{grzedzielski_etal:10a} to study the fate of solar-wind heavy ions in the heliosheath. The He ions are treated as test particles undergoing various binary interactions (BI), with other particle populations constituting the heliosheath plasma. This allows us to calculate (1) how the He ions change their charge states and energies, (2) what the emissivity of He ENA is as a function of energy for each point in the heliosheath, and (3) how many of these ENAs, having survived losses on the way, will reach Earth's vicinity from a particular direction. 

\section{Physical model}
\label{physcialmodel}
In our model we follow the time evolution of charge states and energy of He ions in the heliosheath as solar wind plasma flows from the TS, towards the heliopause (HP), and finally to the heliospheric tail (HT). We treat $\alpha$-particles and He$^+$ ions as test particles, carried by the general flow, which undergo BI with background electrons and protons, with solar ionizing photons, and with neutral H and He atoms coming from interstellar space. In addition we take He ions energy changes into account due to (adiabatic) compression/decompression of the background flow, as well as energy change (in fact decrease) resulting from Coulomb scattering on plasma background and possible effects of spatial diffusion.

\subsection{Evolution of helium ions in phase space}
\label{evolutionof}
To describe the behavior of the two He ion species, $\alpha$-particles and He$^+$ ions, we calculate changes in the local velocity distribution functions $f^{\alpha}$ and $f^{\mathrm{He}^+}$ resulting from displacement of the considered plasma parcel along its flow line determined by the hydrodynamic time-independent solution for background plasma. The background flow is assumed to be stationary in time and axially symmetric, depending on distance $r$ from the Sun and angle $\theta$ from the apex axis (cf. Sect.~\ref{heliosheathbackground}). The functions $f^{\alpha}$ and $f^{\mathrm{He}^+}$ are assumed to be isotropic in velocity space; that is, they are functions of the scalar momentum $p$. Their dependence on $r$ and $\theta$ can be expressed as functions of the curvilinear coordinate $s$ along the flow line or, equivalently, using 
\begin{equation}
 \mathrm{d}s=|v_{\mathrm{sw}}|\mathrm{d}t\, ,
 \label{dsasdt}
\end{equation}
as functions of time $t$, i.e., the flow history of the parcel of background plasma. $v_\mathrm{sw}$ describes the solar wind bulk velocity in the heliosheath. 

The changes in $f^{\alpha}$ and  $f^{\mathrm{He}^+}$ along the flow line are thus determined by coupled equations of the type used to describe the transport of cosmic rays \citep{jokipii:87a}
\begin{eqnarray} \nonumber
v_{\mathrm{sw}}\frac{\mathrm{d}}{\mathrm{d} s}f^{\alpha}&=
\frac{1}{3}\vec{\nabla}\cdot\vec{v}_{\mathrm{sw}}\frac{\partial}{\partial(\ln p)}f^{\alpha}
+G_{\mathrm{BI,He}^+\rightarrow\alpha}-L_{\mathrm{BI,}\alpha\rightarrow\mathrm{He}^+}+\\&-L_{\mathrm{BI,}\alpha\rightarrow\mathrm{He}}
-L_{\mathrm{C,}\nu_{\epsilon}^{\alpha \backslash\mathrm{p}}}-L_{\alpha,\mathrm{H}}+
\vec{\nabla}\cdot\left( \tens{\kappa}\cdot \vec{\nabla} f^{\alpha} \right) \, ,
\label{falpha}
\end{eqnarray}
\begin{eqnarray} \nonumber
v_{\mathrm{sw}}\frac{\mathrm{d}}{\mathrm{d} s}f^{\mathrm{He}^+}&=
\frac{1}{3}\vec{\nabla}\cdot\vec{v}_{\mathrm{sw}}\frac{\partial}{\partial(\ln p)}f^{\mathrm{He}^+}
+G_{\mathrm{BI,}\alpha\rightarrow\mathrm{He}^+}-L_{\mathrm{BI,He}^+\rightarrow\alpha}+\\&-L_{\mathrm{BI,He}^+\rightarrow\mathrm{He}}
-L_{\mathrm{C,}\nu_{\epsilon}^{\mathrm{He}^+ \backslash\mathrm{p}}}-L_{\mathrm{He}^+,\mathrm{H}}+
\vec{\nabla}\cdot\left( \tens{\kappa}\cdot \vec{\nabla} f^{\mathrm{He}^+} \right) \, .
\label{fheplus}
\end{eqnarray}
The successive terms on the righthand side of Eq.~(\ref{falpha}) describe changes in $f^{\alpha}$ due to adiabatic compression/rarefaction of the background flow, gain ($G_{\mathrm{BI,He}^+\rightarrow\alpha}$) from BI conversion of He$^+$ into $\alpha$, loss ($L_{\mathrm{BI,}\alpha\rightarrow\mathrm{He}^+}$) from BI conversion of $\alpha$ into He$^+$, and loss ($L_{\mathrm{BI,}\alpha\rightarrow\mathrm{He}}$)  from BI conversion of $\alpha$ into He, loss ($L_{\mathrm{C,}\nu_{\epsilon}^{\alpha \backslash\mathrm{p}}}$)  due to Coulomb scattering on background protons corresponding to energy loss rate $\nu_{\epsilon}^{\alpha \backslash\mathrm{p}}$ as given by \citet{huba:02a}, and loss ($L_{\alpha,\mathrm{H}}$) due to $\alpha$ interaction with neutral hydrogen \citep{berger_etal:05a}. The last term on the righthand side describes possible effect of $\alpha$-particles spatial diffusion, with corresponding tensorial diffusion coefficient $\tens{\kappa}$ (cf. Sect.~\ref{adiabaticheating}).
Equation~(\ref{fheplus}) is analogous to Eq.~(\ref{falpha}) with symbols $\alpha$ and He$^+$ interchanged. Details on BI are given in Sect.~\ref{binaryinteractions}. 

Equations (\ref{falpha}) and (\ref{fheplus}) do not contain Fokker-Planck type terms that would describe possible local stochastic acceleration. This acceleration is often invoked in the context of ACR populations, though its relevance -- in view of the V1 and V2 data -- is debatable (Florinski et al., 2011, ``The global heliosphere during the recent solar minimum'' talk at the Solar Minimum Workshop, Boulder CO, May 17 -- 19, 2011). Applying stochastic acceleration to the present context of $\sim$~keV ions would require a sound understanding of the small-scale magnetohydrodynamic turbulence in the heliosheath, which is lacking at present.

In numerical solution of Eqs.~(\ref{falpha}) and (\ref{fheplus}) we calculate the evolution of $\alpha$ and He$^+$ spectra separately for each flow line. We set initial discretized spectra at the TS (cf. Sect.~\ref{aparticles}) and then calculate the evolution between adjacent points. Discretization is fixed at 500 bins between 0~km~s$^{-1}$ and 5000~km~s$^{-1}$ with a 10~km~s$^{-1}$ width each. As the plasma parcel proceeds along its flow line, particles are shuffled between bins as required by the interactions. This scheme is repeated for all flow lines.
 
To be able to calculate the righthand side of Eqs.~(\ref{falpha}) and (\ref{fheplus}), one should know, besides the relevant cross sections, the background solar plasma and neutral interstellar gas flows, e.g., the solar wind electron ($n_\mathrm{e}$), proton ($n_{\mathrm{p}}$) densities, the density distribution of interstellar neutral H and He ($n_{\mathrm{H}}$, $n_{\mathrm{He}}$), as well as the corresponding bulk flow velocities of the heliosheath solar wind plasma ($\vec{v}_\mathrm{sw}$), interstellar gas ($\vec{v}_{\mathrm{H}}$), and the effective relative velocities of particles at collisions ($v_\mathrm{rel}$) resulting from local particle velocity distribution functions. Modeling of these functions is described in Sect.~\ref{heliosheathbackground}. 

\subsection{Binary interactions affecting the He ions}
\label{binaryinteractions}
\begin{figure*}
\sidecaption
  \includegraphics[width=12cm]{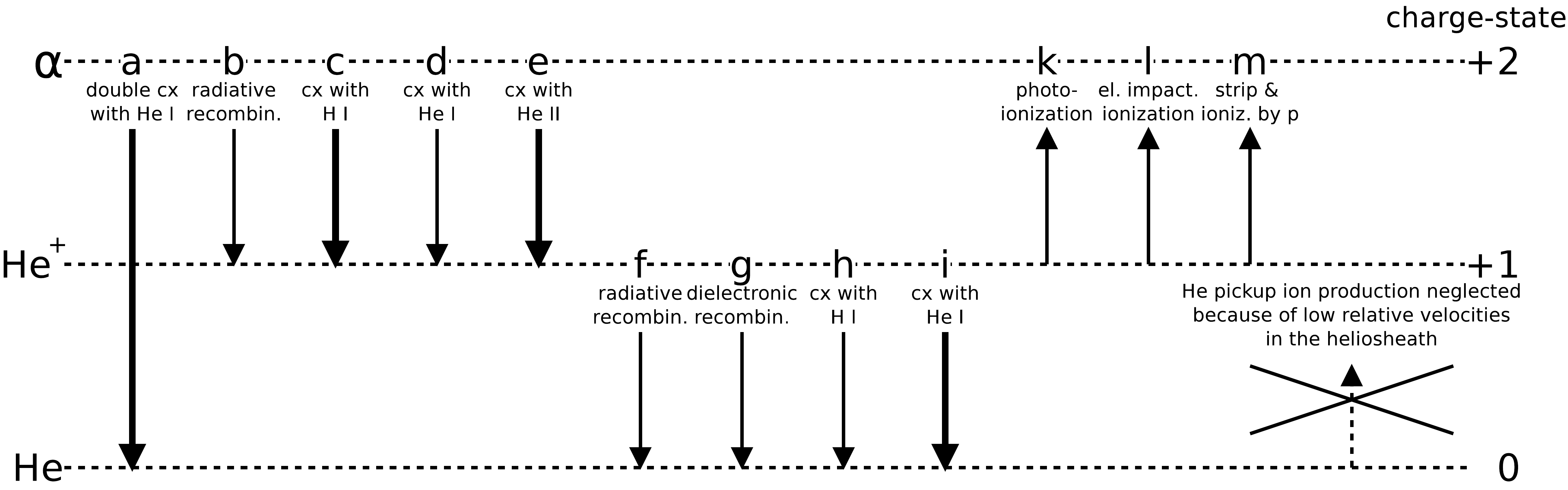}
     \caption{Scheme of transitions between charge states of He resulting from the considered binary interaction with heliosheath background ion, electron, and neutral atom populations (cf. text). The heavy arrows denote transitions of greatest importance in most of heliosheath regions, the dashed arrow (crossed over) is the neglected production of low-energy He$^+$ ions. Emissivity (source funtion) of He ENA is determined by transitions a+f+g+h+i.}
     \label{figbi}
\end{figure*}

We assume that the He ions (test particles) are immersed in a substratum constituted by background heliosheath protons and electrons, background neutral H and He atoms of interstellar origin, and ionizing solar photons. The binary interactions include radiative and dielectronic recombinations, electron impact ionizations, photoionizations, double and single charge exchanges (also to upper levels), and electron stripping. The scheme of transitions between the charge states of He ions resulting from the interactions is shown in Fig.~\ref{figbi}, in which the three levels describe He charge-states 0, +1, +2 and the arrows correspond to binary interactions denoted a, b, c, d, e, f, g, h, i, k, l, m (Table~\ref{tabBI}). 
As we look for He ENAs with velocities $>$~100 km~s$^{-1}$ (energy $>$~207~eV), we take the He$^+$ pickup ions born in the supersonic solar wind into account, while we disregard those originating in the heliosheath, where the relative velocity between heliosheath flow and interstellar neutral He atoms is rather low. 

Interactions b, c, d, e, k, l, m shuffle the ions between charge states 1 and 2. They preserve the total number of ionized He atoms. Interactions a, f, g, h, i  convert He ions into He ENAs and are the source of presumed He ENA fluxes at Earth. Once a He ENA is born, it is assumed to be lost to the heliosheath He budget. Probability of reionization of a 1~keV He atom flying from 5000~AU in the tail to Earth is only 12\%.

\begin{table}
\caption{Binary interactions determining charge-state changes of He ions in the heliosheath}
\label{tabBI}   
\begin{center}                        
\begin{tabular}{c l l}       
\hline\hline                 
Sign & Reaction & Reference \\    
\hline                        
a&	$\alpha$, He $\rightarrow$ He, $\alpha$ (double cx)	& (1) A104\tablefootmark{y}		\\ 
b&	$\alpha$ radiative recombination with e		& (2)			\\ 
c&	$\alpha$, H $\rightarrow$ He$^+$, p			& (3)\tablefootmark{z}	\\
d&	$\alpha$, He $\rightarrow$ He$^+$, He$^+$		& (1) A98, B112		\\                   
e&	$\alpha$, He$^+$ $\rightarrow$ He$^+$, $\alpha$ 	& (1) A84		\\
f&	He$^+$ radiative recombination with e			& (4)			\\ 
g&	He$^+$ dielectronic recombination with e		& (4)			\\ 
h&	He$^+$, H $\rightarrow$ He, p				& (1) A62		\\
i&	He$^+$, He $\rightarrow$ He, He$^+$			& (1) A70		\\
k&	photoionization of He$^+$				& (5)			\\ 
l&	electron impact ionization of He$^+$ 			& (6)			\\
m&	stripping/cx ionization by p				& (1) A48, A54		\\
\hline                                   
\end{tabular}
\end{center}
\tablefoot{e -- electron, p -- proton, $\alpha$ -- $\alpha$-particle, cx -- charge exchange\\
\tablefoottext{y}{A104 and similars symbols denote reactions as listed in (1)}\\
\tablefoottext{z}{(1) A88 for energy $>$ 5 keV, excitation and/or emission involving upper levels includes (1) B90, B96, B102, B104}}
\tablebib{
(1)~\citet{redbooks}; 
(2)~\citet{arnaud_rothenflug:85a}; 
(3)~\citet{liu_etal:03a}; 
(4)~\citet{aldrovandi_pequignot:73a}; 
(5)~\citet{bochsler_etal:12b}; 
(6)~\citet{janev_etal:87a}
}
\end{table}

\subsection{Heliosheath background plasma}
\label{heliosheathbackground}

The state of bulk solar wind plasma in the heliosheath is, at present, the subject of an intense debate brought about by unexpected results of the plasma experiment on board Voyager-2 (V2) \citep{richardson_etal:08a} during and after crossing(s) of the TS in Aug./Sep. 2007. There is little doubt that the downwind (post-TS) plasma is in a very different state from what was expected on the basis of standard Rankine-Hugoniot equations. The post-shock temperature of the majority (thermal) protons seems to be  much lower ($\sim$70\,000~K) than expected in a single-fluid shock transition ($\sim$10$^6$ K), and the bulk flow velocity starts to decrease well ahead of the shock with a much smaller velocity jump at the shock itself. Also the electron temperature seems to be quite low, $T_{\mathrm{e}}<10$~eV \citep{richardson:08a}. 

In contrast, a relatively small fraction ($\sim$10-30\%) of post-shock protons is endowed with energies of $\sim$1 to perhaps several keV. This nonthermal proton population, resulting presumably from ionized interstellar H atoms picked up by the supersonic solar wind, is thought to contain the bulk of total energy density (pressure) of the post-shock plasma. V2 data indicate such a situation seems to prevail deep into the post-shock plasma \citep{richardson:11a} with little spatial change observed in thermal proton density and temperature. Such conditions may exist in the frontal lobes and perhaps also in near tail of the heliosheath. 
Another important fact is a much faster than anticipated decline in heliosheath plasma bulk velocity derived from Voyager-1 (V1) data as the spacecraft receded from the TS in years 2005-2011 \citep{krimigis_etal:11a}. In particular the simultaneous drastic decrease on Aug. 25, 2012 in the fluxes of ions of energies $>0.5$~MeV/n observed by the LET telescopes of the cosmic ray subsystem instrument on board V1 and the accompanying increase in the magnetic field to about 0.41~nT (ftp://lepvgr.gsfc.nasa.gov/pub/voyager/) seem to indicate that V1 might have crossed the heliopause around this date.

These results imply that the TS-HP stretch along V1 trajectory is only $\sim121.5 - 94  = \sim 27.5$~AU long; that is, the heliosheath is much narrower than $\sim$60 -- $\sim$100~AU obtained in gasdynamical modeling based on data available prior to V1 and V2 TS-crossings (e.g., in the \citet{izmodenov_alexashov:03a} model the distances Sun-TS and Sun-HP for ecliptic latitudes corresponding to the V2 trajectory are 110 AU and 208 AU, respectively). A narrow heliosheath of only $25\pm8$~AU thickness in the upwind region also comes out from the analysis of H ENA fluxes observed by the IBEX, SOHO/HSTOF, and Cassini/INCA spacecraft \citep{hsieh_etal:10a}. 

These new findings are at present not properly integrated and understood within a coherent physical picture. Despite the success of a realistic and time-varying description of the TS crossing positions by both Voyager spacecraft in a recent model by \citet{washimi_etal:11a}, the single-fluid MHD calculations employed therein are unable to render the nonthermal aspects of the particle distribution functions in the heliosheath that may be important for He ions physics. Therefore to numerically describe the background plasma flow conditions we employ three time-independent, axisymmetric heliosheath models (denoted in the following: hydrodynamic, Parker, and ad hoc), corresponding to three simple variants of assumed heliosheath plasma.

The hydrodynamic model is the model developed by \citet{izmodenov_alexashov:03a} and used previously in \citet{grzedzielski_etal:10a}. The Parker and ad-hoc models, though not internally coherent in terms of physics, are `tailored' in such a way as to approximately render the V2 suggested spatial distribution of variables decisive for the behavior of He ions: that is, thermal proton and electron densities ($n_{\mathrm{p}}$, $n_{\mathrm{e}}$), temperatures ($T_{\mathrm{p}}$, $T_{\mathrm{e}}$), as well as the main traits of the nonthermal plasma components. We also try to approximately render the common heliosheath bulk flow velocity ($v_{\mathrm{sw}}$) along the V2 trajectory. For each variant, integration of Eqs.~(\ref{falpha}) and (\ref{fheplus}) is performed. In this way we follow the time evolution of the charge states, spatial distribution, and energy spectra of He ions in each fluid element carried by the background flow. 
This allows the local He ENA production rates to be calculated and -- after accounting for He ENA energy losses and reionization on the way -- we can construct expected He ENA spectra at Earth. The point is to test how sensitive the predicted He ENA fluxes are to the assumed widely discordant variants of plasma. We believe that if important common features of predicted He ENA fluxes appear in all considered cases, credence could be given to the results despite the partial inadequacy of the physical modeling employed.

{\em In the hydrodynamic model}, the background flow of solar  plasma and neutral hydrogen atoms in the supersonic solar wind, inner heliosheath, and distant heliospheric tail is described as single-fluid, non-magnetic, gas-dynamical flow of heliospheric proton-electron plasma coupled by mass, momentum, and energy exchange with the neutral interstellar hydrogen atoms calculated kinetically (Monte-Carlo approach). The Sun as a source of solar wind and ionizing photons is assumed to be spherically symmetric, with the wind speed of 450~km~s$^{-1}$, Mach number 10, and $n_\mathrm{p}= 7\ \mathrm{cm}^{-3}$ at Earth orbit. At infinity, a uniform interstellar flow of 25~km~s$^{-1}$ with neutral hydrogen density $n_{\mathrm{H,LISM}}= 0.2\ \mathrm{cm}^{-3}$, proton density $n_{\mathrm{p,LISM}}= 0.07\ \mathrm{cm}^{-3}$, and temperature 6000~K was assumed. 

Interstellar neutral He atoms are represented by a uniform substratum with density $n_{\mathrm{He}}= 0.015\ \mathrm{at.\,cm}^{-3}$ \citep{gloeckler_etal:04a} flowing with velocity of 25~km~s$^{-1}$. Because of axial symmetry, all variables depend on the radial distance $r$ from the Sun and angle $\theta$ from the apex direction. The background flow is found in form of $n_{\mathrm{p}}$, $n_{\mathrm{e}}$, $T_{\mathrm{p}}$ ($=T_{\mathrm{e}}$) given as functions of distance $s$ along individual flow lines. There are 180 flow lines, each starting at Earth's orbit. The flow lines are identified by the initial (1~AU) value of the angle $\theta$. In this solution the TS and HP are, correspondingly, 102 and 177~AU distant from the Sun along the apex axis. 

The shocked solar wind plasma is very hot immediately behind the TS ($T_{\mathrm{p}}$ = $T_{\mathrm{e}}$ $\sim$10$^6$~K) and then cools down to about 20\,000~K in the distant heliotail. The background plasma velocity distributions are always local Maxwellians.

\begin{figure*}
\centering
   \includegraphics[width=0.32\textwidth]{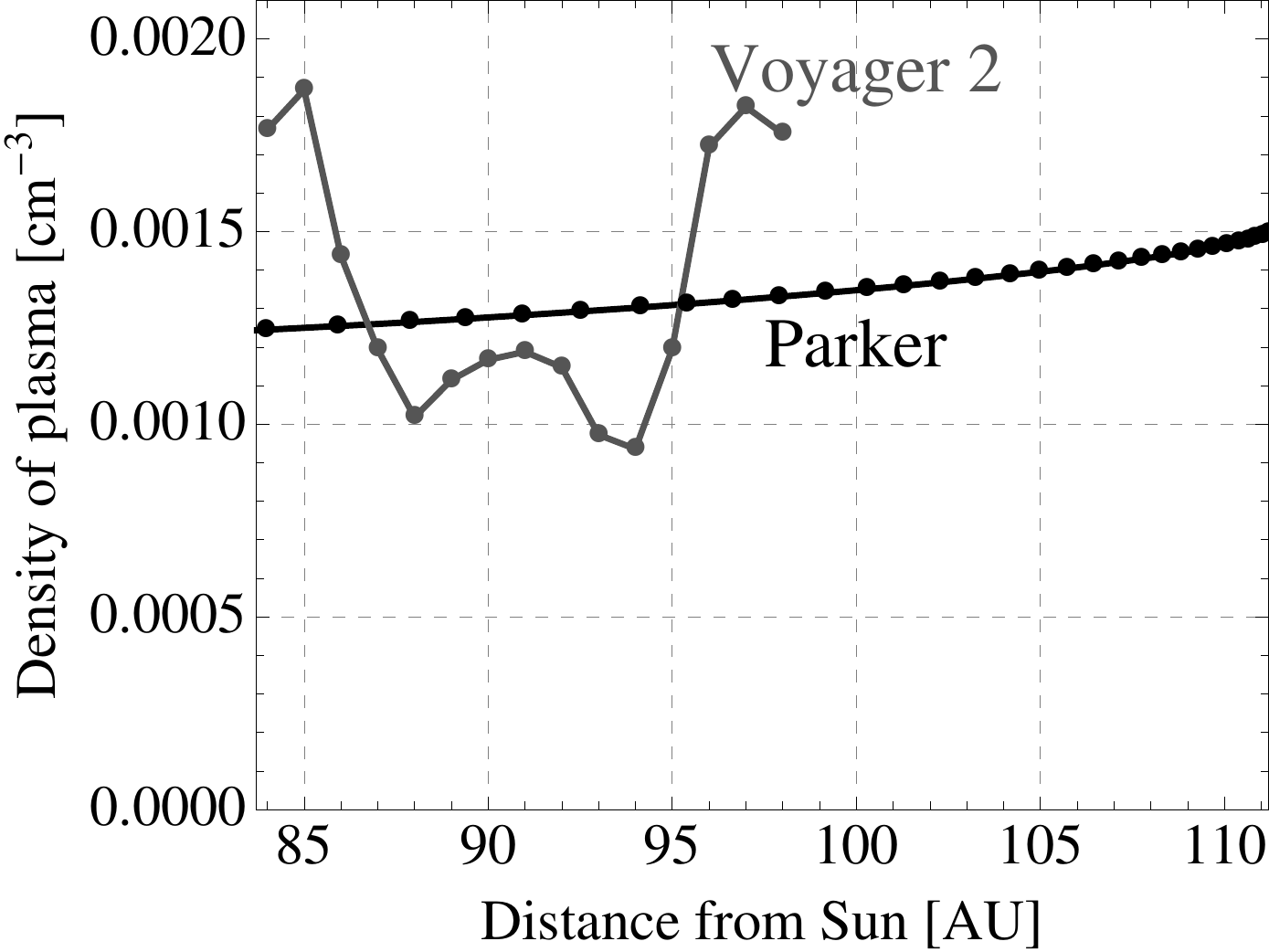}
   \includegraphics[width=0.32\textwidth]{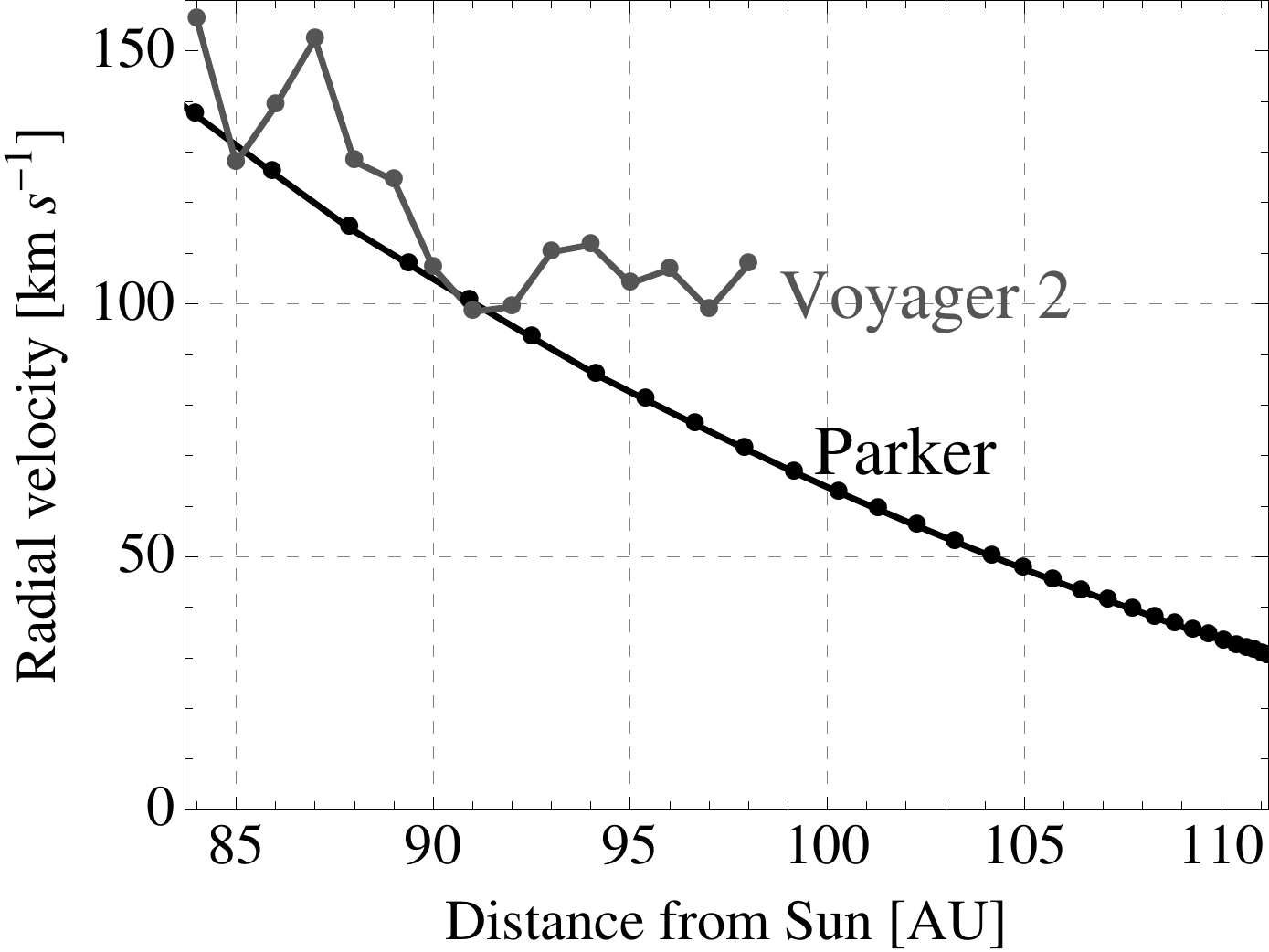}
   \includegraphics[width=0.32\textwidth]{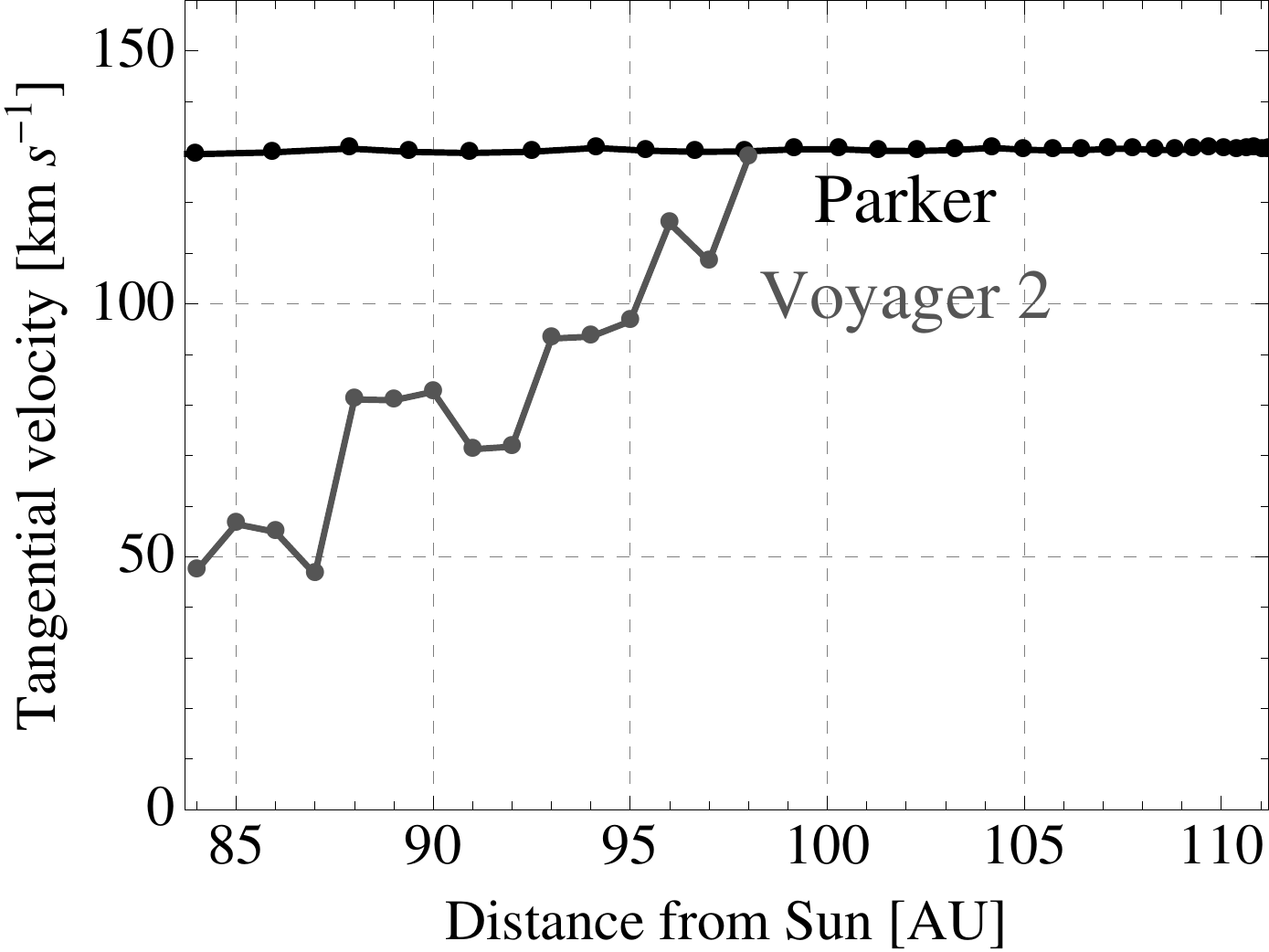}
   \caption{Comparison of heliosheath proton density (left panel), radial velocity (middle panel), and tangential velocity (right panel) in the Parker model with corresponding in situ measurements \citep{richardson:11a} along the orbit of Voyager 2.}
   \label{figv2vsparker}
\end{figure*}

{\em The Parker model} is the classical subsonic solution for a point source of incompressible fluid (stellar wind) in a uniform, incompressible external flow \citep{parker:61a}. To be applicable, the model requires the TS radius much less than the distance Sun-HP. We assume that the solar wind mass source and the interstellar flow are the same as in the hydrodynamical model. The background velocities and shape of the flow lines are determined by the Parker analytical solution. Typical heliosheath velocities are in the range 120--250~km~s$^{-1}$ at $\sim$100~AU and tend to 170~km~s$^{-1}$ in the tail. We crudely approximate observed heliosheath plasma conditions by assuming constant heliosheath proton density to be constituted by two proton populations, the thermal one with $n_{\mathrm{p,th}}= 0.0015\ \mathrm{cm}^{-3}$ and another nonthermal one with $n_{\mathrm{p,nth}}= 0.0005\ \mathrm{cm}^{-3}$. 
The proton energies are described by a Maxwellian with 70\,000~K temperature and a monoenergetic population of 1.1~keV per proton, respectively. 
Electron temperature is set to $T_{\mathrm{e}}= 3\ \mathrm{eV}$. To conform to a smaller HP distance as presently envisaged \citep{krimigis_etal:11a} the Sun-HP stretch along the apex axis is set to 101~AU. The corresponding Sun-HP distance at the V2 ecliptic latitude is $83.7+27.5=111.2\ \mathrm{AU}$. Comparison of proton density, bulk radial, and tangentional velocities measured in the heliosheath by V2 with the Parker model solution is shown in Fig.~\ref{figv2vsparker}.

{\em The ad hoc model} was developed for this research. We include in it both the new (smaller) heliospheric scale and some of the new physical aspects of heliosheath plasmas. The solar wind mass source and the interstellar flow are again the same as in the hydrodynamic model. However, the background flow in the heliosheath is solved anew. We use the approximation that the flow along each of the flow lines is one-dimensional in a known channel of varying cross section $A(s)$. The geometry of the channels is determined by the flow lines of the hydrodynamic model but linearly rescaled so as to place the termination shock (with shape conserved) at the observed distance of the V2 crossing, that is, at 83.7 AU from the Sun. In this rescaling, the heliopause is put at $111.2\ \mathrm{AU}$ (at V2 ecliptic latitude). The flow along the reshaped flow lines is found by integrating equations of conservation of mass, momentum, and energy with the appropriate source terms on the 
righthand side. 

In a channel we have the mass conservation equation
\begin{equation}
\frac{1}{A}\frac{\textrm{d}}{\textrm{d} s}\left[A\rho_\mathrm{p}v_{\mathrm{sw}}\right]=Q \, ,
\end{equation}
where $\rho_{\mathrm{p}}$ is total (thermal + nonthermal) mass density of protons and $v_{\mathrm{sw}}$ bulk plasma velocity in the heliosheath, and $Q$ is the mass-loading term. We neglect the contribution of solar wind He ions and neutral interstellar He atoms to mass loading compared to the H contribution, and we also neglect electron impact ionization of H atoms because of low heliosheath electron temperature, $T_{\mathrm{e}}=3$~eV. Mass loading results only from proton -- H-atom charge exchange and net $Q=0$.

\begin{figure*}
\centering
   \includegraphics[width=0.32\textwidth]{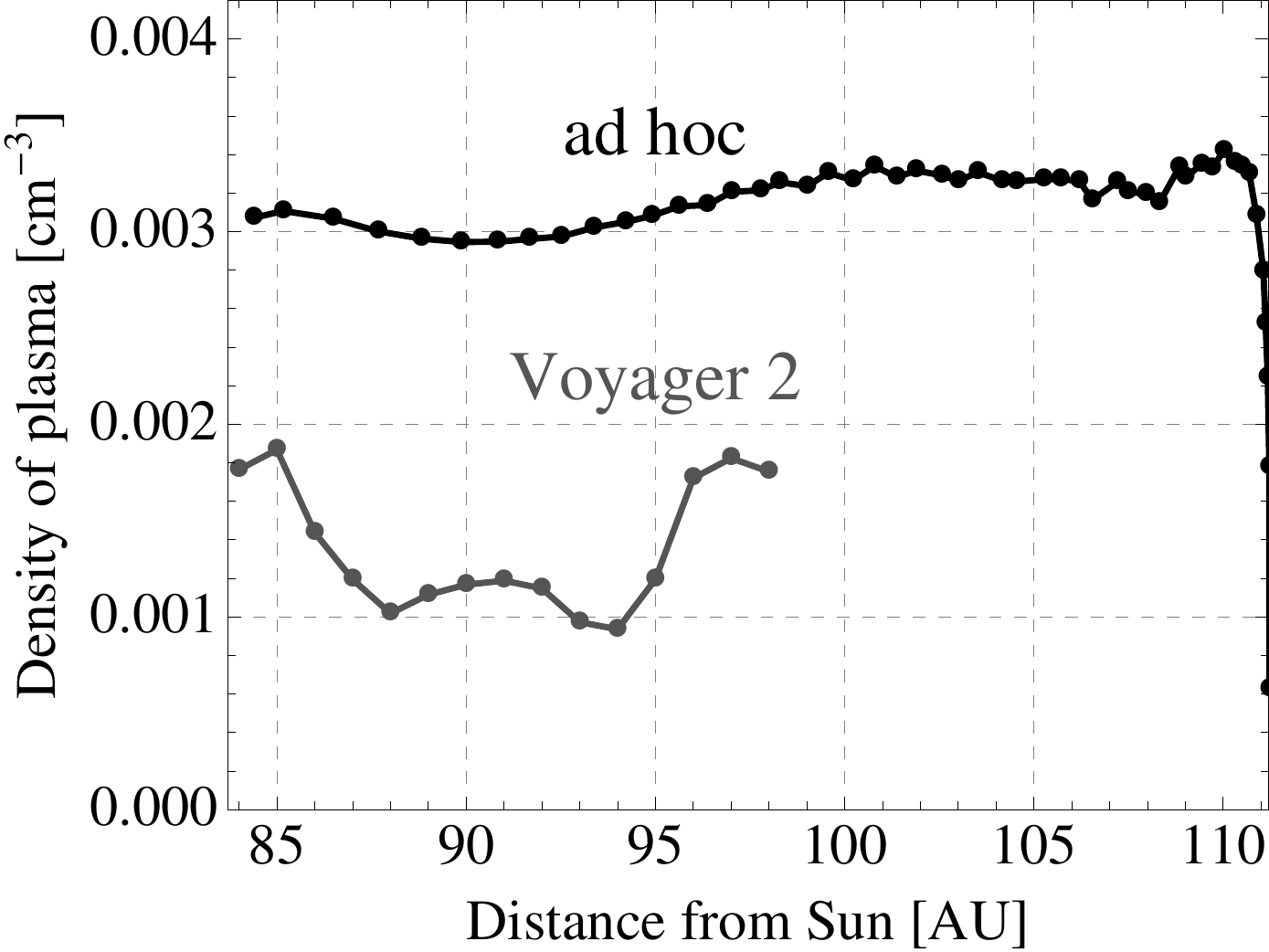}
   \includegraphics[width=0.32\textwidth]{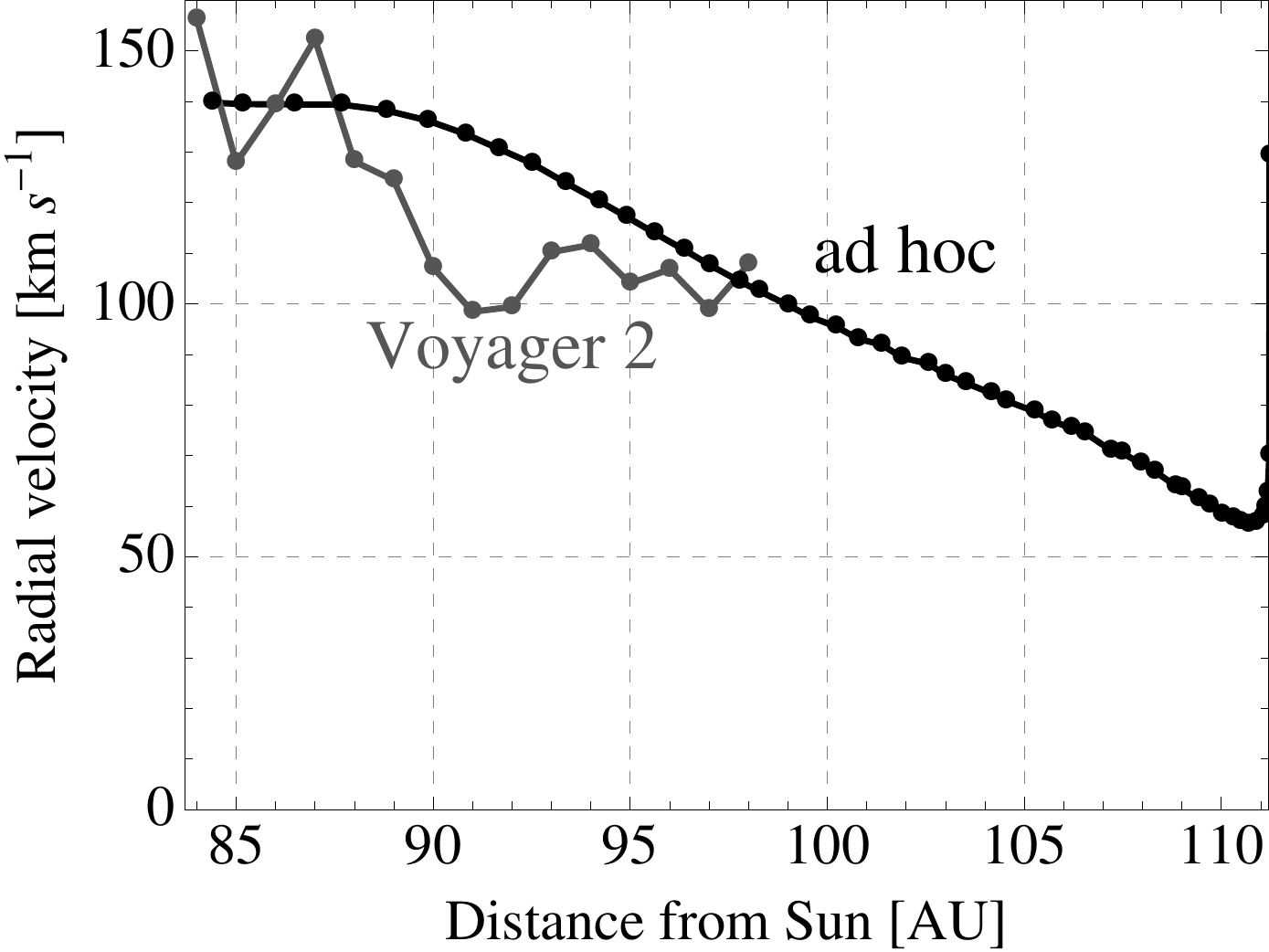}
   \includegraphics[width=0.32\textwidth]{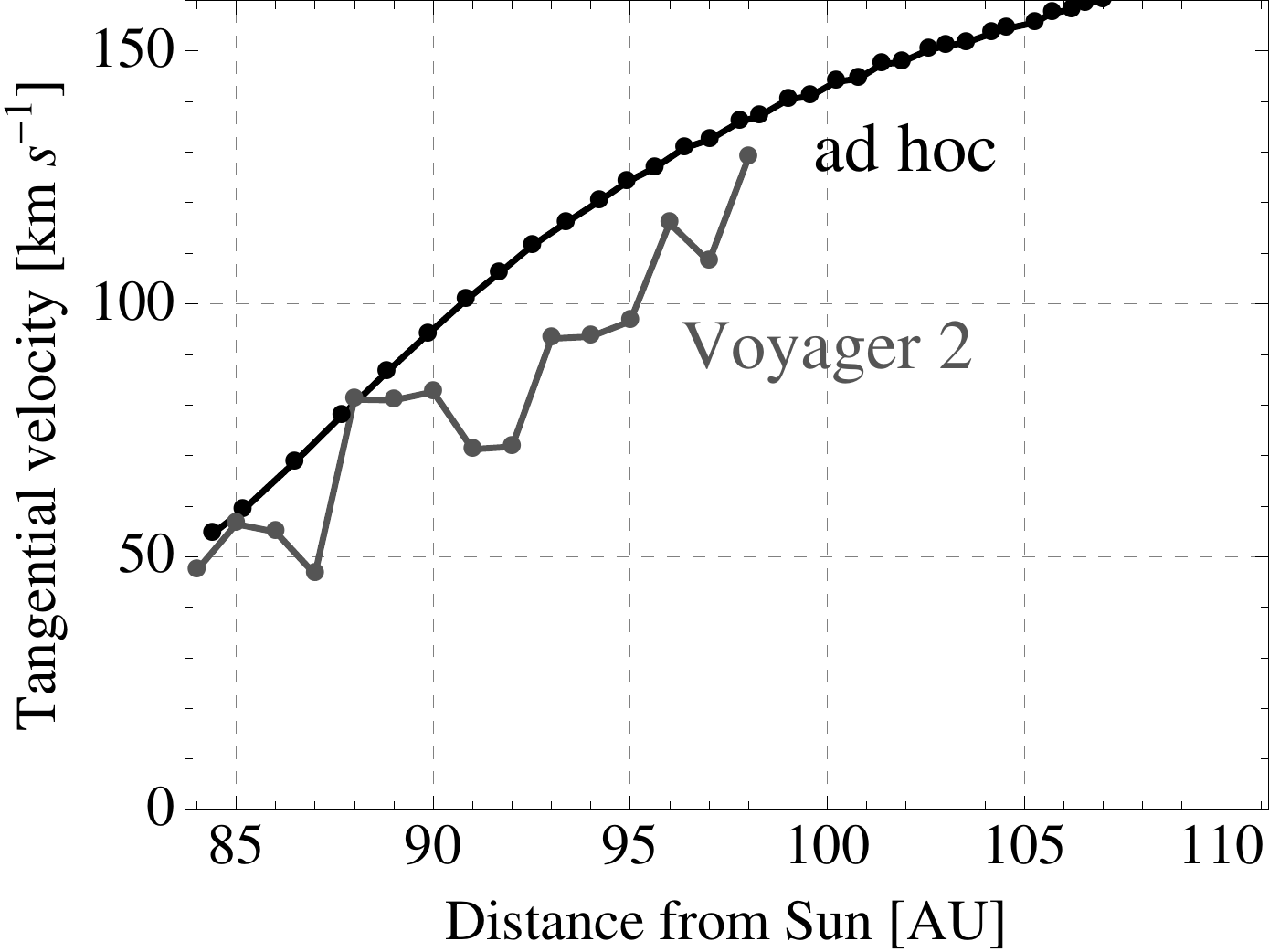}
   \caption{Comparison of heliosheath proton density (left panel), radial velocity (middle panel), and tangential velocity (right panel) in the ad hoc model with corresponding in situ measurements \citep{richardson:11a} along the orbit of Voyager 2.}
   \label{figv2vsadhoc}
\end{figure*}

Momentum loading is nonvanishing, and we have in the equation of motion the frictional force (acting in direction $\vec{s}$) resulting from proton -- H-atom charge exchange:
\begin{equation}
\frac{1}{A}\frac{\textrm{d}}{\textrm{d} s} \left[ A\rho_{\mathrm{p}}v_{\mathrm{sw}}^2\right]=-\frac{\textrm{d}}{\textrm{d}s}p_{\mathrm{nth}}+
 \sigma(v_{\mathrm{eff}})v_{\mathrm{eff}}n_\mathrm{H}\rho_{\mathrm{p}}\left(v_\mathrm{H}\cos \theta-v_{\mathrm{sw}}\right) \, ,
\end{equation}
where $\sigma$ denotes the charge exchange cross section, and $v_{\mathrm{eff}}$ is the relative velocity between particles. In the energy equation we treat as small the kinetic energy of the bulk heliosheath flow, as well as the pressure of the thermal protons (kept at $T_{\mathrm{p,th}}=70\,000\ \mathrm{K}$), and retain solely the terms proportional to the high pressure (energy) $p_\mathrm{nth}$, of the mono-energetic nonthermal proton population. We assume this energy (1) is convected with flow velocity $v_\mathrm{sw}$ and (2) decays by charge exchange with interstellar H atoms on a time $\tau_{\mathrm{cx}}$ depending on particle energy. Then the equation takes the form
\begin{equation}
\frac{1}{A}\frac{\textrm{d}}{\textrm{d} s}\left[ Av_{\mathrm{sw}}\gamma p_{\mathrm{nth}} \right]= (\gamma-1)v_{\mathrm{sw}}\frac{\textrm{d}}{\textrm{d} s}p_{\mathrm{nth}}-\frac{p_{\mathrm{nth}}}{\tau_{\mathrm{cx}}}\, ,
\end{equation}
where $\gamma$ is the usual adiabatic exponent. In numerical calculations we assume the nonthermal population contains 25\% of mass and average energy per proton at the TS is 1.1~keV \citep{giacalone_decker:10a}. This means that nonthermal protons contain 77\% of the total energy associated with particle random motions. The bulk velocity on the TS downwind side is 150~km~s$^{-1}$ \citep{richardson_etal:08a} and the TS downwind density dependence on angle $\theta$ is like in the hydrodynamic model but rescaled to the new TS position. The fit of the ad hoc model solution to V2 measurements in the heliosheath for proton density, radial, and tangential velocities is shown in Fig.~\ref{figv2vsadhoc}.

The density in the ad hoc model seems to be too high by a factor $\sim$2. A number of causes could be responsible, among them time-dependent effects in the solar wind flow (our modeling always uses an `average' solar wind model) and/or heliosheath asymmetry resulting from a skewed interstellar magnetic field.

\subsection{Adiabatic heating/cooling versus spatial diffusion}
\label{adiabaticheating}

Changes in background plasma density along the flow line induce He ion energy changes that can be treated as adiabatic if diffusion is slow enough. We made an estimate of the possible role of diffusion in the case of the hydrodynamic model. To find regions in the heliosheath in which diffusion of $\alpha$-particles is negligible we use a simplified version of Eq.~(\ref{falpha}) treated as the transport equation for pressure $p_\alpha$ of the $\alpha$-particle cosmic ray gas \citep{drury_voelk:81a}. In this approximation Eq.~(\ref{falpha}) becomes
\begin{equation}
 \frac{\mathrm{d}}{\mathrm{d}t} \left( \frac{p_\alpha}{\gamma-1} \right)=\vec{\nabla}\cdot\left[\kappa \vec{\nabla}\left(\frac{p_\alpha}{\gamma-1}\right) \right]+\frac{\gamma}{\gamma-1}\frac{p_\alpha}{\rho}\frac{\mathrm{d}\rho}{\mathrm{d}t}\, ,
 \label{eqdiff}
\end{equation}
in which $\rho$ is the background plasma density. We assume diffusion is negligible as long as the first term on the righthand side (effect of diffusion) is much less than the second term (adiabatic heating/cooling). The heliosheath diffusion coefficient for low-energy $\alpha$-particles is in fact unknown. It depends most probably on heliosheath turbulence. Models for the superposition of slab and 2-D turbulence were developed for the solar wind \citep{zank_etal:04a}, but it is unclear how the assumed (2:8) energy partition between the two modes of turbulence corresponds to real heliosheath conditions. Therefore we take simply two limiting formulae as crude guesses: (i) Bohm diffusion, $\kappa_\mathrm{B}$,  and (ii) phenomenological diffusion by \citet{leroux_etal:96a} originally developed for the ACR ions, $\kappa_{\mathrm{ACR}}$. We extrapolate them to our $\alpha$-particle energies. In both formulae we assume $\kappa$ is scalar. 

\begin{figure*}
\sidecaption
  \includegraphics[width=6cm]{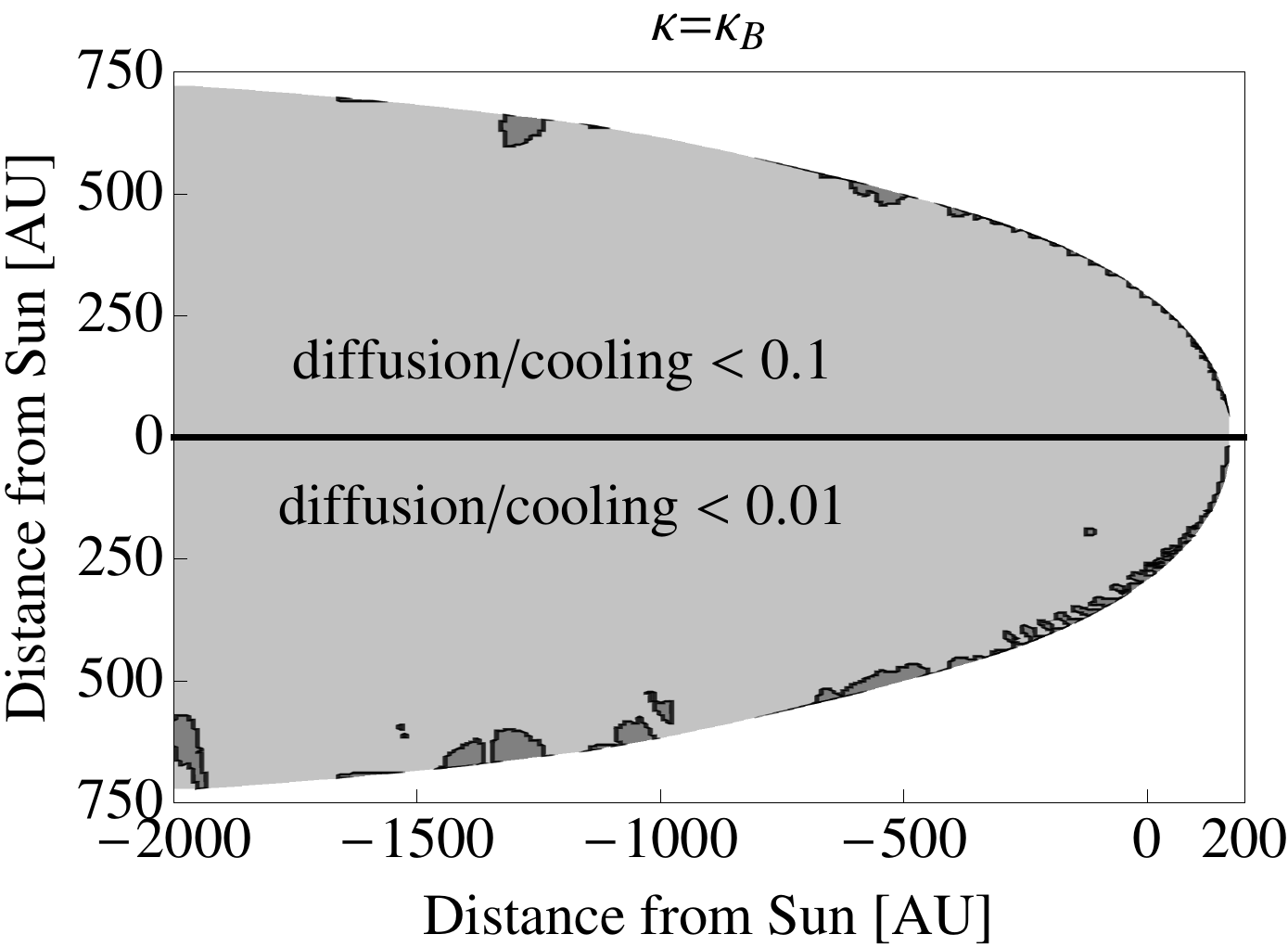}
  \includegraphics[width=6cm]{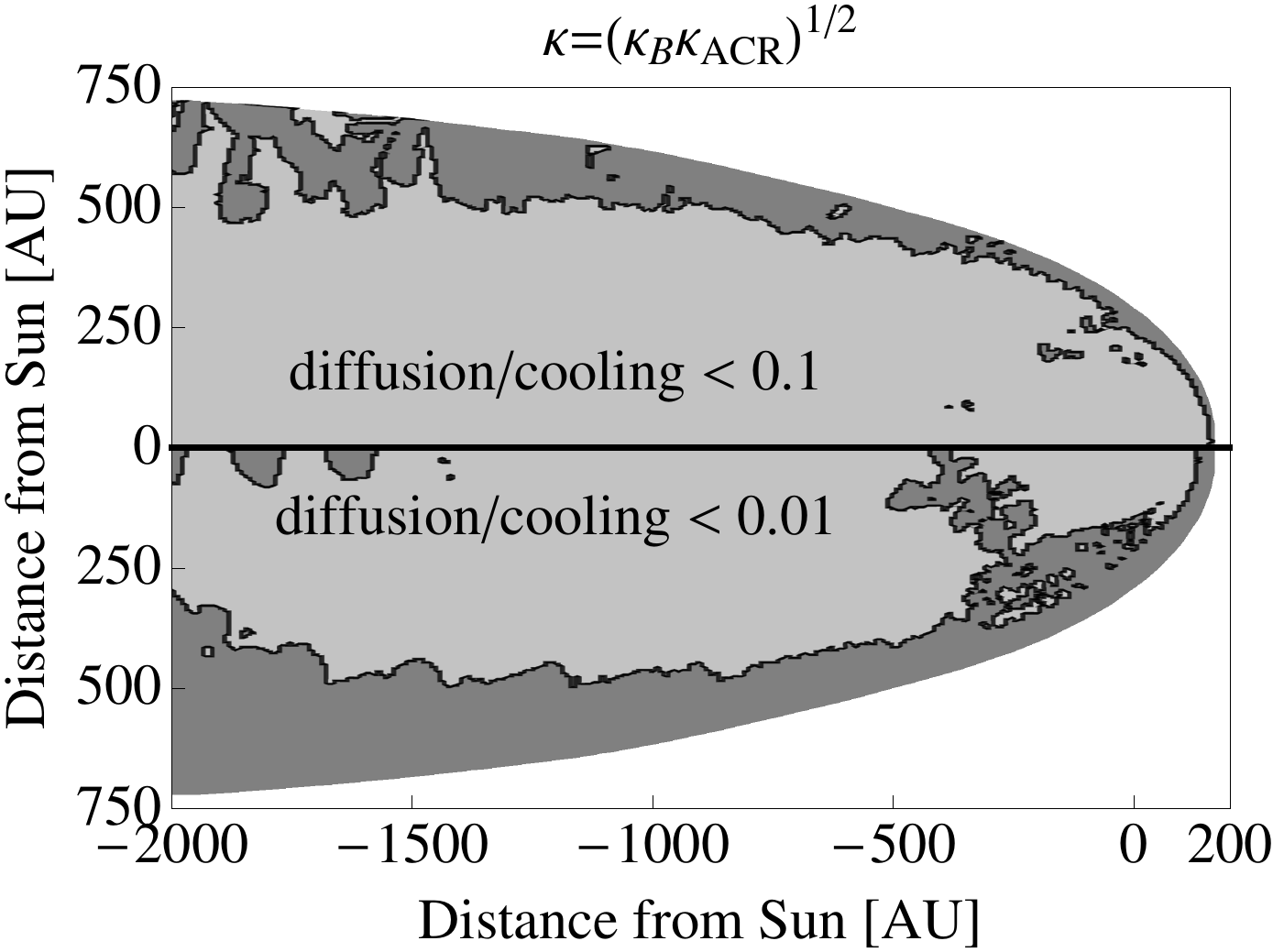}
   \caption{Heliosheath map of regions (light-gray areas) where in the hydrodynamic model the ratio of the 1st-to-2nd terms on the righthand side of. Eq.~(\ref{eqdiff}) is $<1/100$ (lower panels) and $<1/10$ (upper panels); that is, the role of diffusion is small. Left column is for Bohm diffusion, case (i); right column for $(\kappa_{\mathrm{B}}\kappa_{\mathrm{ACR}})^{1/2}$, case(iii). }
   \label{figdiff}
\end{figure*}

Because diffusion coefficients (i) and (ii) differ, for the range of energies discussed ($\sim$0.2 -- 20 keV, cf. Sect.~\ref{expectedHe}) by 4.5 to 5.5 orders of magnitudes, we also take for comparison (iii) an intermediary diffusion coefficient equal to $(\kappa_{\mathrm{B}}\kappa_{\mathrm{ACR}})^{1/2}$. In calculating the values of the first term on the righthand side of Eq.~(\ref{eqdiff}) along the flow line, one has to use the local value of the heliosheath magnetic field $B_\mathrm{hsh}$. As the detailed structure of $B_{\mathrm{hsh}}$ is unknown, we calculate $B_\mathrm{hsh}$ along the flow line assuming the field starts on the downwind side of the TS with a value of 0.1~nT as measured by \citet{burlaga_etal:05a,burlaga_etal:08a} (variation in the magnetic field along the TS surface was described by standard dependence on angle/distance). 
Further evolution of the field then followed from background plasma density variations. In this we assumed that the magnetic field is frozen to background plasma, its direction is randomly oriented and plasma compression/decompression is isotropic, i.e. $B_{\mathrm{hsh}}\propto(\mathrm{background\,plasma\,density})^{2/3}$. 
In this way one can construct maps of the ratio of first-to-second terms on the righthand side of Eq.~(\ref{eqdiff}) in the heliosheath. The maps are shown for values of the said ratio equal to 1/100 and 1/10 in Fig.~\ref{figdiff}. They correspond to $\kappa=\kappa_{\mathrm{B}}$ and $\kappa=(\kappa_{\mathrm{B}}\kappa_{\mathrm{ACR}})^{1/2}$ and to $\alpha$-particle energies at injection, peaking at energy corresponding to 170~km~s$^{-1}$ (cf. Sect.~\ref{aparticles}). Figure~\ref{figdiff} suggests that neglect of diffusion is justified for both diffusion versions (i) and (iii). For version (ii) diffusion is a paramount effect, but this case is hardly realistic for the low energies discussed here. Diffusion effects obtained for He$^+$ ions are similar. 

In the Parker model diffusion does not exist because of assumed uniform background plasma density. In the ad hoc model, the density distribution is quite flat (cf. Fig.~\ref{figv2vsadhoc}, left panel), so diffusion is also negligible. In solving Eqs.~(\ref{falpha}) and (\ref{fheplus}) we therefore always consistently  retain the adiabatic cooling/heating term (first on the righthand side). Energy variations due to these terms are moderate; for instance, energy increase along the flow line starting at $\theta=5\degr$ at the TS from apex, does not exceed a factor 1.5 at its maximum.

\subsection{Density distributions and energy spectra of $\alpha$-particles and He$^+$ ions in the heliosheath}
\label{aparticles}

The $\alpha$-particles ($n_\alpha$) and He$^+$ ions ($n_{\mathrm{He}^+}$) density distributions and energy spectra are calculated for the three background flow models (hydrodynamic, Parker, ad hoc, Sect.~\ref{heliosheathbackground}) by integrating Eqs.~(\ref{falpha}) and (\ref{fheplus}) along 180 flow lines. The approximations and procedures we adopted are described in Sects.~\ref{binaryinteractions}, \ref{heliosheathbackground} and \ref{adiabaticheating}. Consistent treatment applies effectively to the interval (100, 2000)~km~s$^{-1}$ corresponding to energy range 0.207 --~82.9~keV for He ions. The initial values of $n_\alpha$ and $n_{\mathrm{He}^+}$ are stated at the TS for the hydrodynamic and ad hoc models, assuming that the total He content of the solar wind constitutes 5\% of the local proton plasma (by number). For the Parker model they are stated at the distance fitted to TS crossing by V2.

Two variants of injected (initial) $\alpha$-particle energy distributions at the TS were taken: one peaking at energy 0.6~keV, which corresponds to velocity 170~km~s$^{-1}$ (in the following we label this variant ``170~km~s$^{-1}$''), the other peaking at energy 2.1~keV, which corresponds to 320~km~s$^{-1}$ (we label this variant ``320 km~s$^{-1}$''). These energies were chosen because the first one could correspond to the case when excess post-TS energy of an $\alpha$-particle is coming from the bulk velocity jump at the TS, and the second to the case when excess energy is approximately equal to total bulk kinetic energy at the shock \citep{richardson_etal:08a} (cf. Sects.~\ref{introduction} and \ref{heliosheathbackground}). In both variants the initial velocity distribution shape is the same: a kappa-distribution with $\kappa = 2.5$. 
This value was chosen because it corresponds to the borderline between the near-equilibrium region applicable to heavy solar wind ions and the far-equilibrium region corresponding to the inner heliosheath \citep[cf. their Fig.~2]{livadiotis_mccomas:11a}. 

The initial population of He$^+$ at the TS is made up of (a) solar wind $\alpha$-particles singly-decharged during their flight to the TS and (b) pickup He$^+$ produced by neutral interstellar He ionization in the supersonic solar wind region. Based on solar wind as in the hydrodynamic model, we estimated that group (a) amounts to 0.0005 (by number) of the proton content. For group (b) the corresponding number is 0.002 \citep{rucinski_etal:03a}. Initial spectra for He$^+$ group (a) at the TS are the same as those for $\alpha$-particles. For He$^+$ group (b) we consistently use only one initial spectrum: a kappa(=2.5)-distribution that peaks at 4~keV.  

\begin{figure*}
\centering$
\begin{array}{c c c}
   \includegraphics[width=0.4161\textwidth]{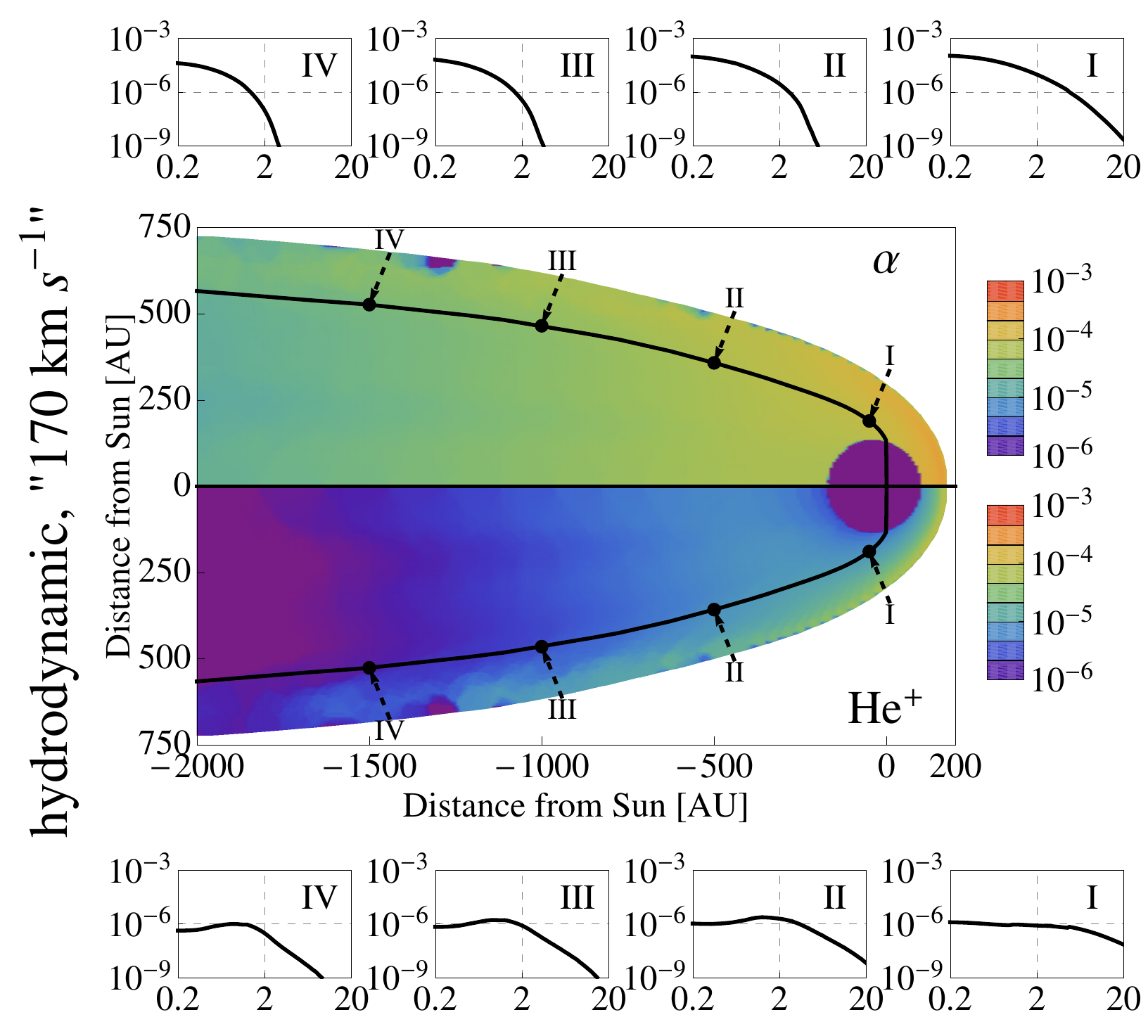} &
   \,\,\,\,\,\,\,\,\,\,\,\,\,\,\,\,\,\,\,\,\,&
   \includegraphics[width=0.4161\textwidth]{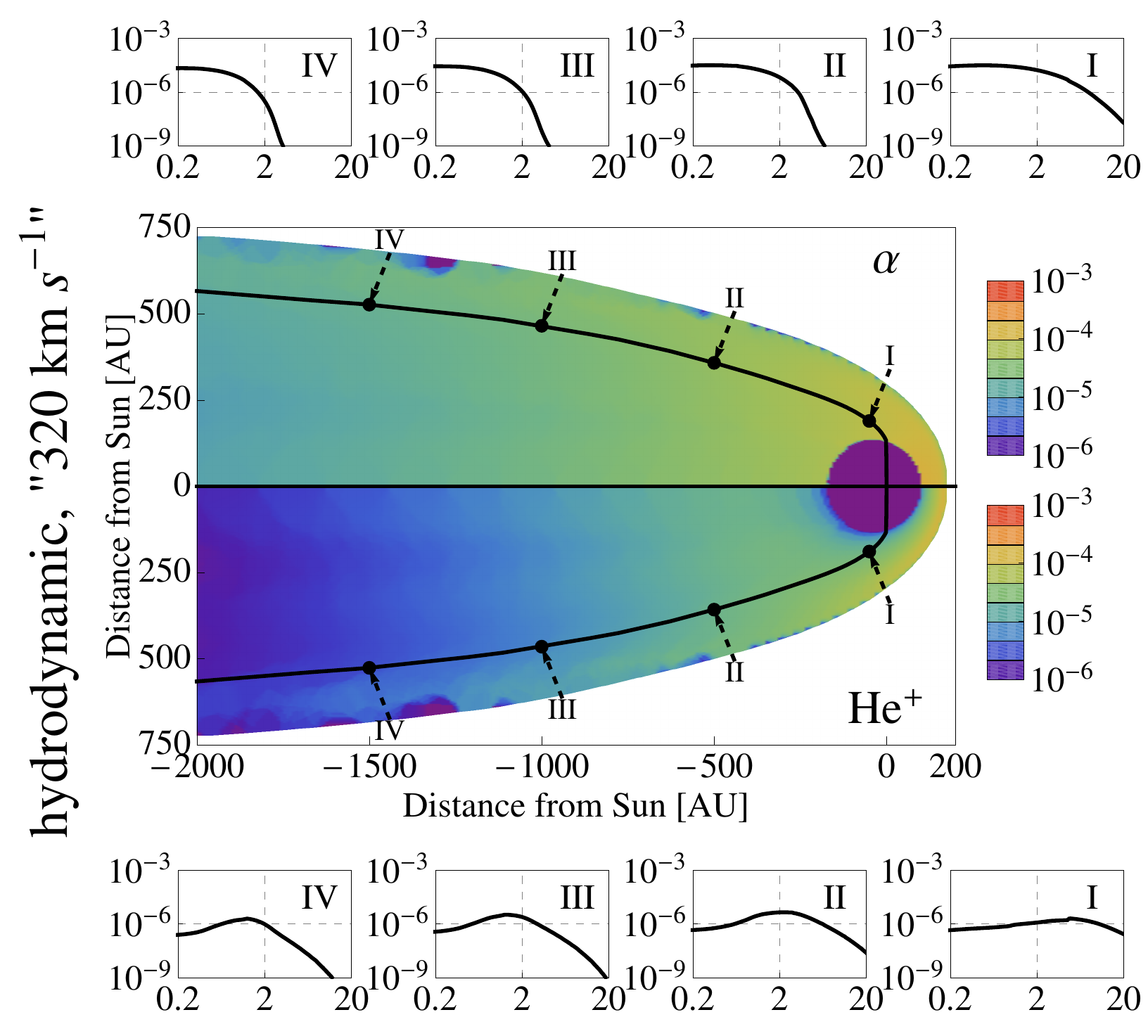} \\
   
   \includegraphics[width=0.4161\textwidth]{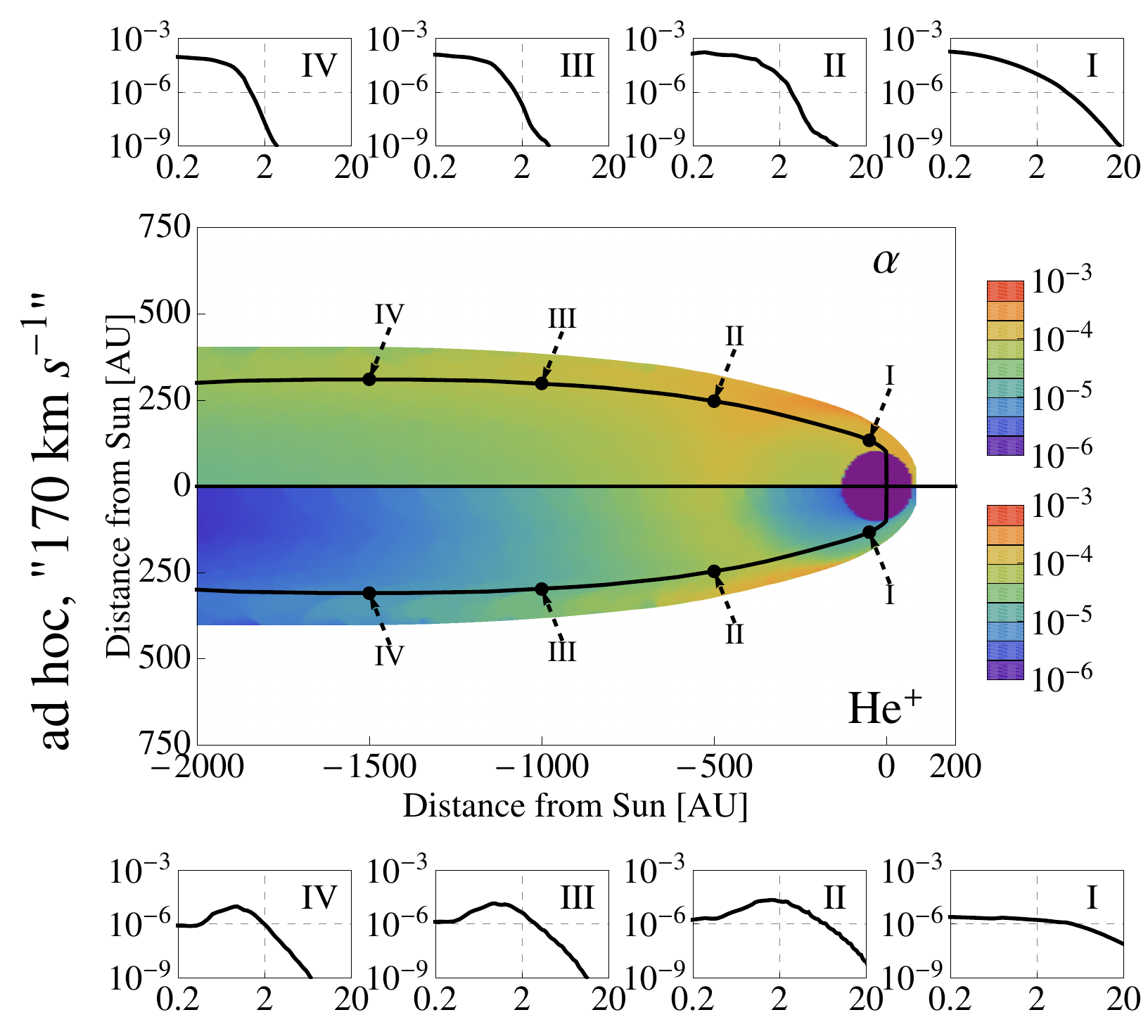} &
   &
   \includegraphics[width=0.4161\textwidth]{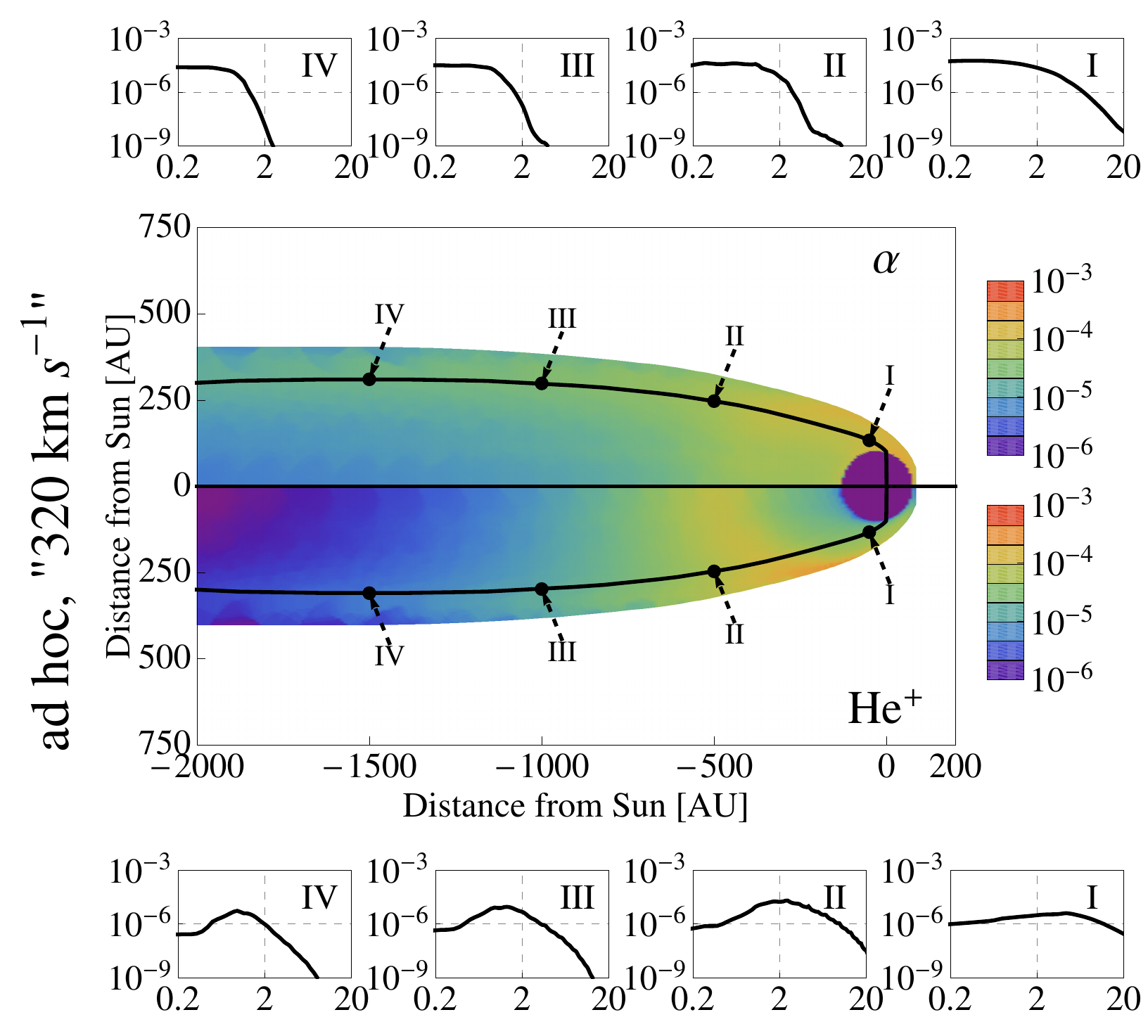} \\
   
   \includegraphics[width=0.4161\textwidth]{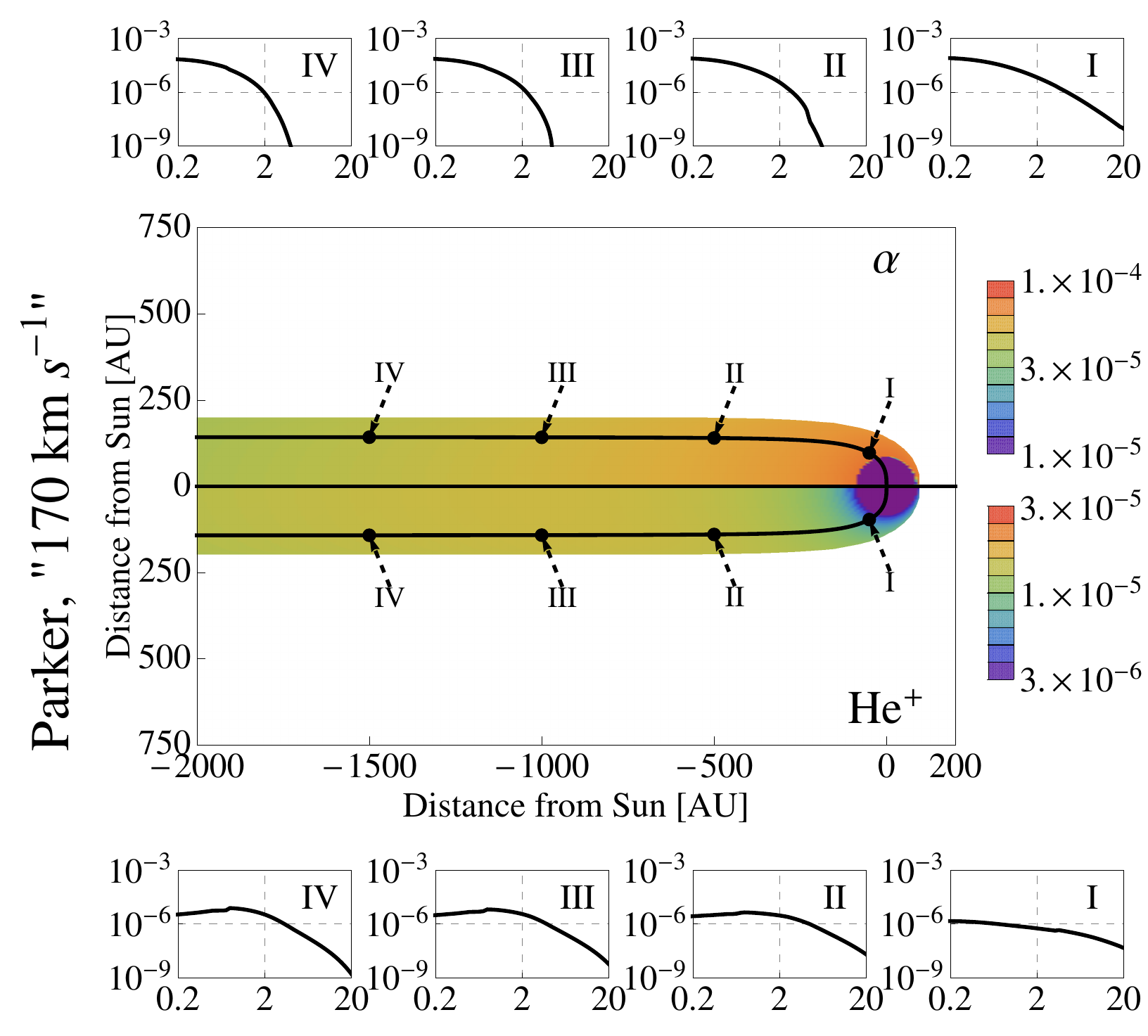} &
   &
   \includegraphics[width=0.4161\textwidth]{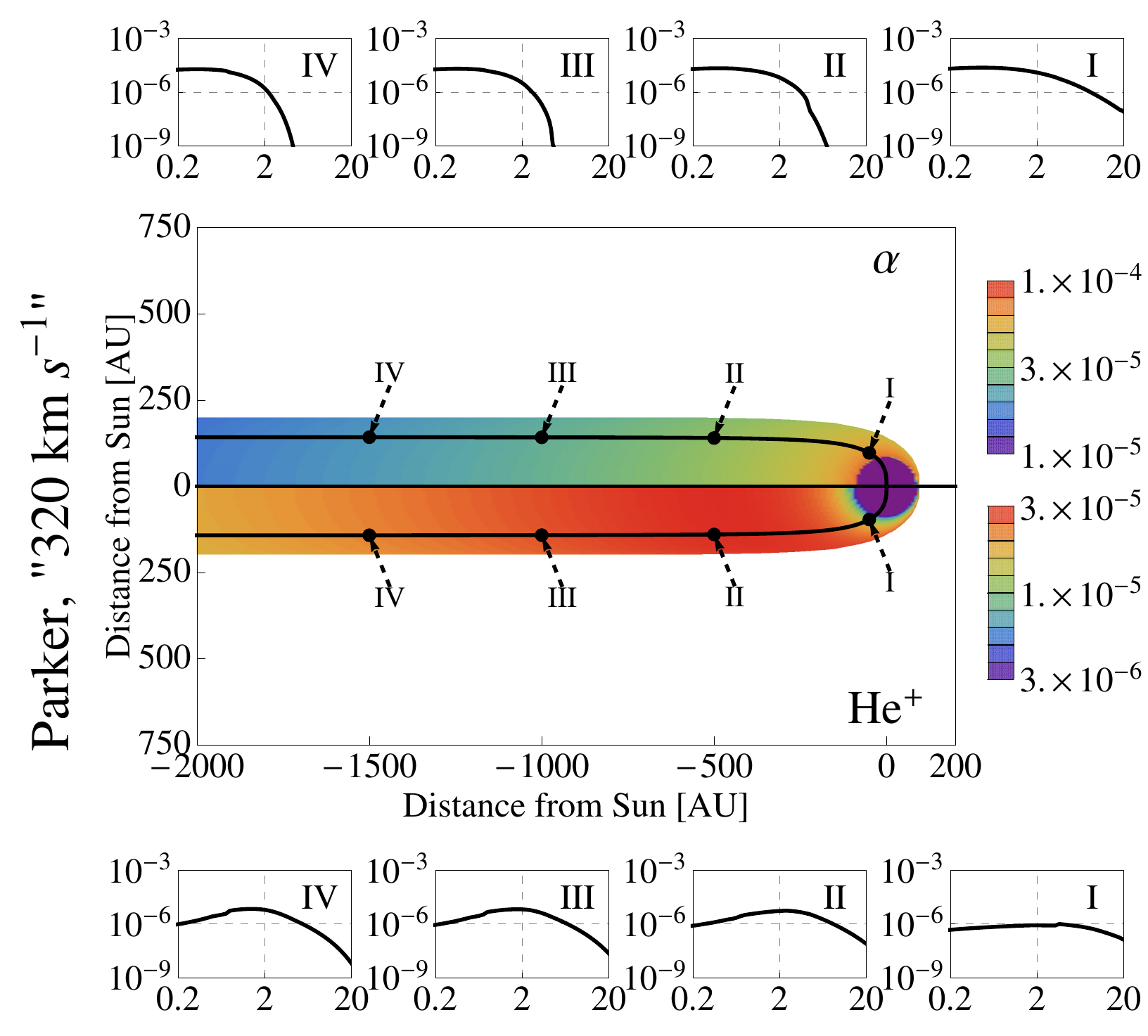} \\
\end{array}$
   \caption{Calculated density distributions and energy spectra of heliosheath energetic He ions for the hydrodynamic, Parker, and ad hoc models. Upper/middle/lower rows correspond to hydrodynamic/ad hoc/Parker models. Left column of the panels corresponds to the peak velocity of the initial kappa-distributions of $\alpha$-particles injected at TS equal to 170 km~s$^{-1}$; right column, correspondingly, to 320 km~s$^{-1}$. He$^+$ injection as described in the text. In a given row, the upper (lower) half of each heliosphere map represents density distribution of $\alpha$-particles (He$^+$ ions). Distance scale is in AU.
   Density scales (equal for $\alpha$-particles and He$^+$ ions only in the two upper rows) are indicated on vertical strips on the right side of each half-heliosphere map. Small panels placed above (below) maps illustrate the evolution of  energy spectra of $\alpha$-particles (He$^+$ ions) as the background plasma parcel is carried over points I, II, III, IV along the flow line. The line starting at the TS point $90\degr$ away from the apex is shown. Horizontal (vertical) axis in small panels describes ion energy in keV (ion density in cm$^{-3}$keV$^{-1}$).}
   \label{figiondensity}
\end{figure*}

Energetic He ion heliosheath density distributions and energy spectra obtained from integration of Eqs.~(\ref{falpha}) and (\ref{fheplus}) are shown in Fig.~\ref{figiondensity} for variants ``170~km~s$^{-1}$'' and ``320 km~s$^{-1}$''.  The lefthand column of panels corresponds to ``170~km~s$^{-1}$â'', the righthand one to ``320~km~s$^{-1}$''. Each row corresponds to one of the three models employed (top to bottom: hydrodynamic, ad hoc, Parker). Upper (lower) half of each heliosphere map refers to $\alpha$-particles (He$^+$ ions). To give some feeling of the evolution of energy spectra, above and below each map we show four panels with ion energy spectra corresponding to spatial positions labeled I, II, III, IV. As indicated, all these points lie on the same flow line starting at the TS at a point $90\degr$ away from the apex as seen from the Sun. 

Figure~\ref{figiondensity} illustrates some of the basic physics of He-ion behavior. First densities per energy interval are much lower for He$^+$ than for $\alpha$-particles. This reflects the differences in the injection at the TS. However, second, while the $\alpha$-particles shift to lower energies when we go from point I to IV with monotonically decreasing spectra preserved, the He$^+$ ions often develop maxima at mid-energy range, even though they decline at high energy. 

This has two main causes. One is that, on the whole, energetic He ions lose energy on average because of Coulomb scattering on the plasma background. Though adiabatic compression may overturn this process, this happens only locally in the upwind heliosheath. In the end, both Coulomb scattering and general decompression towards the tail have to take over. The other cause is related to the circumstance that conversion of $\alpha$s into He$^+$ is, in practice, a one-way process. This comes from the high ionization potential of He$^+$. Therefore while locally heated $\alpha$s may later resupply the mid energies of the He$^+$ population, no similar process operates in the opposite direction. 

\begin{figure*}
\centering$
\begin{array}{c c c}
   \includegraphics[width=0.4167\textwidth]{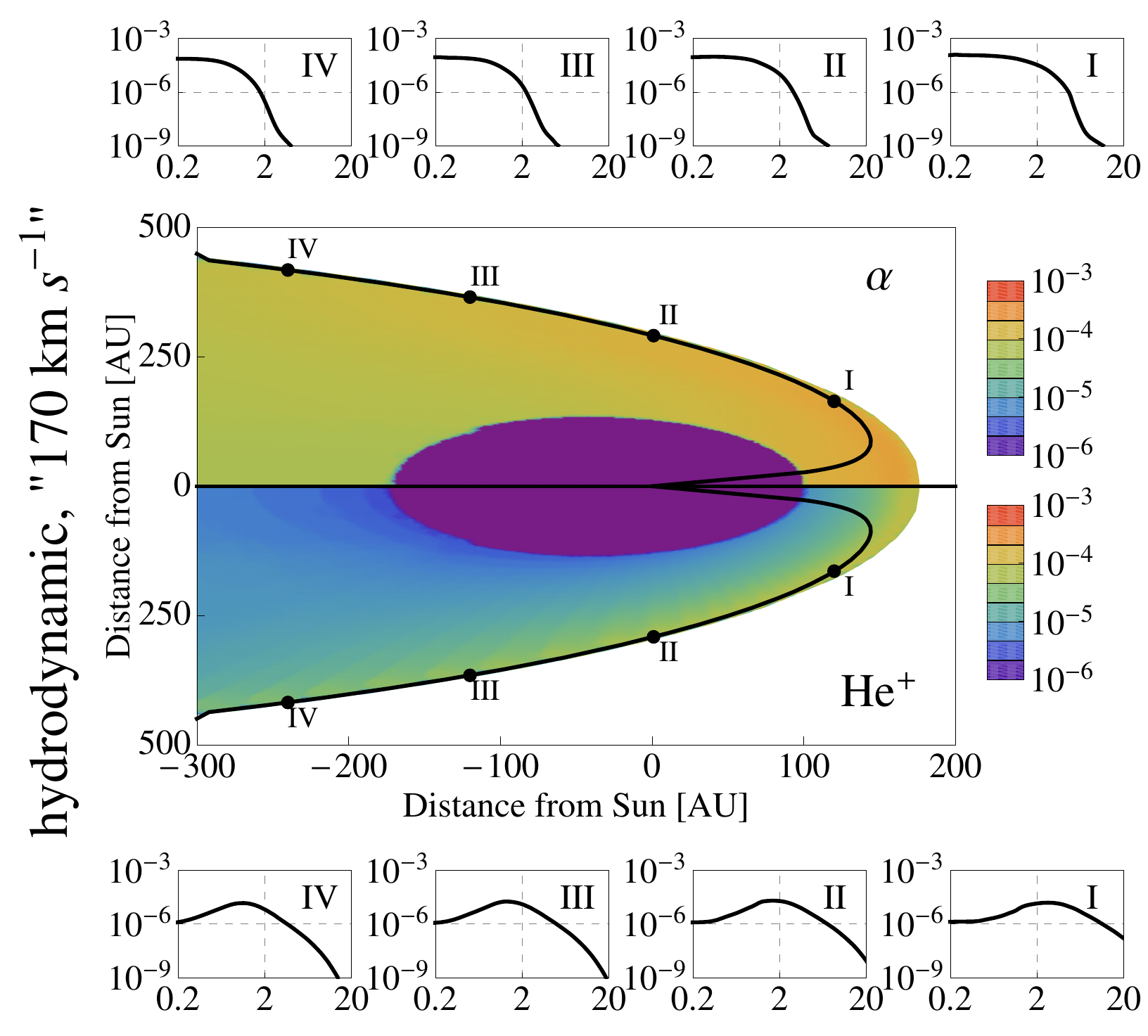} &
   \,\,\,\,\,\,\,\,\,\,\,\,\,\,\,\,\,\,\,\,\,&
   \includegraphics[width=0.4167\textwidth]{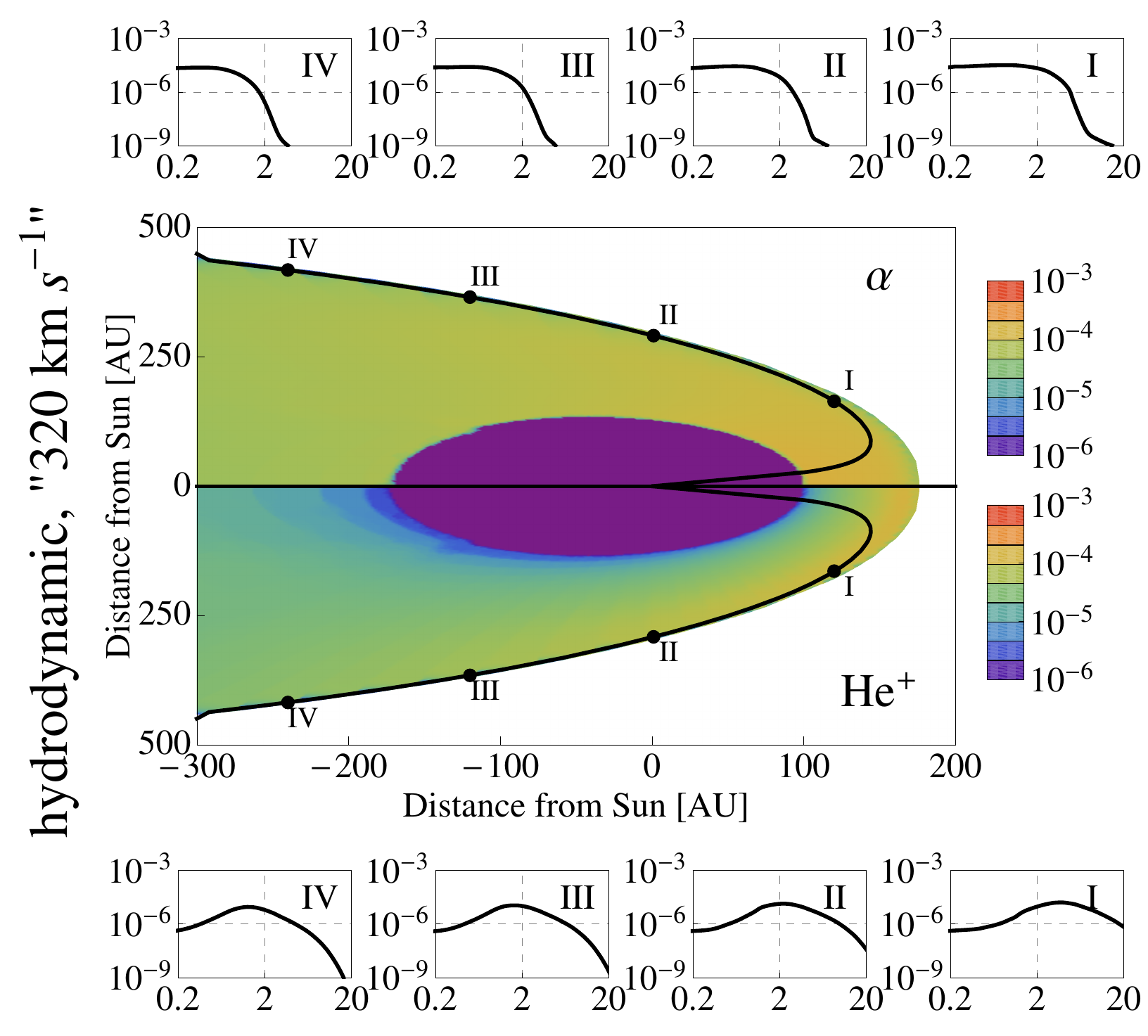} \\
\end{array}$
   \caption{Calculated heliosheath energetic He-ion density distributions and energy spectra for the frontal part of heliosphere (in the hydrodynamic model). Injection data in left panel correspond to ``170 km~s$^{-1}$'', right panel to ``320 km~s$^{-1}$''. Other details similar to Fig.~\ref{figiondensity}, except that the flow line shown starts at a TS point $15\degr$ away from the apex.}
   \label{figiondensityshort}
\end{figure*}

Insight into the situation in the upwind heliosheath for the hydrodynamic model can be gathered from Fig.~\ref{figiondensityshort}, which shows expanded maps of the frontal part of heliosphere in a similar format to Fig.~\ref{figiondensity}. The points I to IV are now shown along a flow line starting at a TS point $15\degr$ away from the apex. The energy maxima of the He$^+$ ions are more pronounced here than in Fig.~\ref{figiondensity}. This is due to more effective heating by compression in the frontal heliosheath.

\section{Expected He ENA fluxes from the heliosheath}
\label{expectedHe}

The local emissivity (source function) $j_\mathrm{ENA}$ (cm$^{-3}$s$^{-1}$keV$^{-1}$) of He ENA is determined by the product of reagents' densities and reaction rates (i.e. relative velocity times cross section, cm$^{3}$s$^{-1}$) for BI a, f, g, h, i (Sect.~\ref{binaryinteractions}). All these quantities can be found as functions of position in the three heliosheath models (hydrodynamic, Parker, ad hoc) by procedures described in Sects.~\ref{evolutionof}, \ref{binaryinteractions}, \ref{heliosheathbackground}, \ref{adiabaticheating}, and \ref{aparticles}. The He ENA intensity $I_{\mathrm{ENA}}(E,\theta)$~(cm$^{-2}$s$^{-1}$sr$^{-1}$keV$^{-1}$) of particles of energy $E$ coming from the direction at angle $\theta$ from the apex is expressed as a line-of-sight (LOS) integral of the source function from the TS to HP
\begin{eqnarray}\nonumber
 I_{\mathrm{ENA}}(E,\theta)=
 L_{\mathrm{sw}}\int_{r_{\mathrm{TS}}}^{r_{\mathrm{HP}}}
 &\frac{j_{\mathrm{ENA}}(E,r,\theta)}{4\pi}&
 F_{\mathrm{CG}}(E,\theta,\vec{v}_{\mathrm{sw}})\times\\
 &&\exp\left[ -\int_{r_\mathrm{TS}}^{r} \frac{\tau_{\mathrm{ext}}(r')}{v_{\mathrm{ENA}}} \mathrm{d}r'\right]
 \mathrm{d}r \, .
 \label{Iena}
\end{eqnarray}

Formula (\ref{Iena}) contains corrections for re-ionization losses in the heliosheath and in the supersonic solar wind, as well as a correction ($F_{\mathrm{­CG}}$) for the Compton-Getting effect. The losses in the heliosheath were calculated for BI as listed in Table~\ref{tabLoss}. In Eq.~(\ref{Iena}) they are described by the effective depth $\tau_{\mathrm{ext}}$ for extinction of He ENA with velocity $v_{\mathrm{ENA}}$. The main contributions to heliosheath losses come from reactions a, i, and s. In the hydrodynamic model, He ENA reionization by electron impact is also of some importance in view of the high electron temperature. The ionization losses in the supersonic solar wind ($L_{\mathrm{sw}}$) are mainly due to reaction p (photoionization) in Table~\ref{tabLoss}, with a small contribution from reaction q inside the orbit of Mars \citep{bzowski_etal:12a}. On the whole, losses inside the supersonic solar wind are much less important than those in the heliosheath.  

The value of $F_{\mathrm{­CG}}$ depends on He ENA energy $E$ in the observer's frame and the velocity of the source carried by the background relative to the Sun ($v_\mathrm{sw}$) for a given model. $F_{\mathrm{­CG}}$ is a function of angular distance $\theta$ of the LOS from apex. It is calculated for an observer at rest, at 1~AU from the Sun in the direction of incoming ENA, with solar gravity neglected. $F_{\mathrm{CG}}$ is particularly significant in reducing energy of He ENA coming from the upwind heliosheath in the ad hoc model, in which background flow recedes quickly from the Sun.

\begin{table}
\caption{Binary interactions  determining He ENA losses in the heliosheath}
\label{tabLoss}   
\begin{center}                        
\begin{tabular}{c l l}       
\hline\hline                 
Sign & Reaction & Reference\\    
\hline    
a\tablefootmark{z}&	$\alpha$, He $\rightarrow$ He, $\alpha$ (double cx) 	& (1) A104					\\
d\tablefootmark{z}&	$\alpha$, He $\rightarrow$ He$^+$, He$^+$		& (1) A98					\\                    
i\tablefootmark{z}&	He$^+$, He $\rightarrow$ He, He$^+$			& (1) A70					\\
p		&	He photoionization					& (2)						\\
q		&	He electron impact ionization				& (3)						\\
r		&	He, p $\rightarrow$ He$^+$, e, p			& (1)  D24     					\\
s		&	He, p $\rightarrow$ He$^+$, H				& (1)  A32 					\\
t		&	He, H $\rightarrow$ He$^+$, H$^-$			& (1)  A10					\\

\hline                                   
\end{tabular}
\end{center}
\tablefoot{e -- electron, p -- proton, H -- hydrogen atom, $\alpha$ -- $\alpha$-particle, cx -- charge exchange;\\
\tablefoottext{z}{same as in Table~\ref{tabBI}}
}
\tablebib{
(1)~\citet{redbooks}; 
(2)~\citet{cummings_etal:02a}; 
(3)~\citet{janev_etal:87a}
}
\end{table}

\begin{figure*}
\centering
   \includegraphics[width=0.48\textwidth]{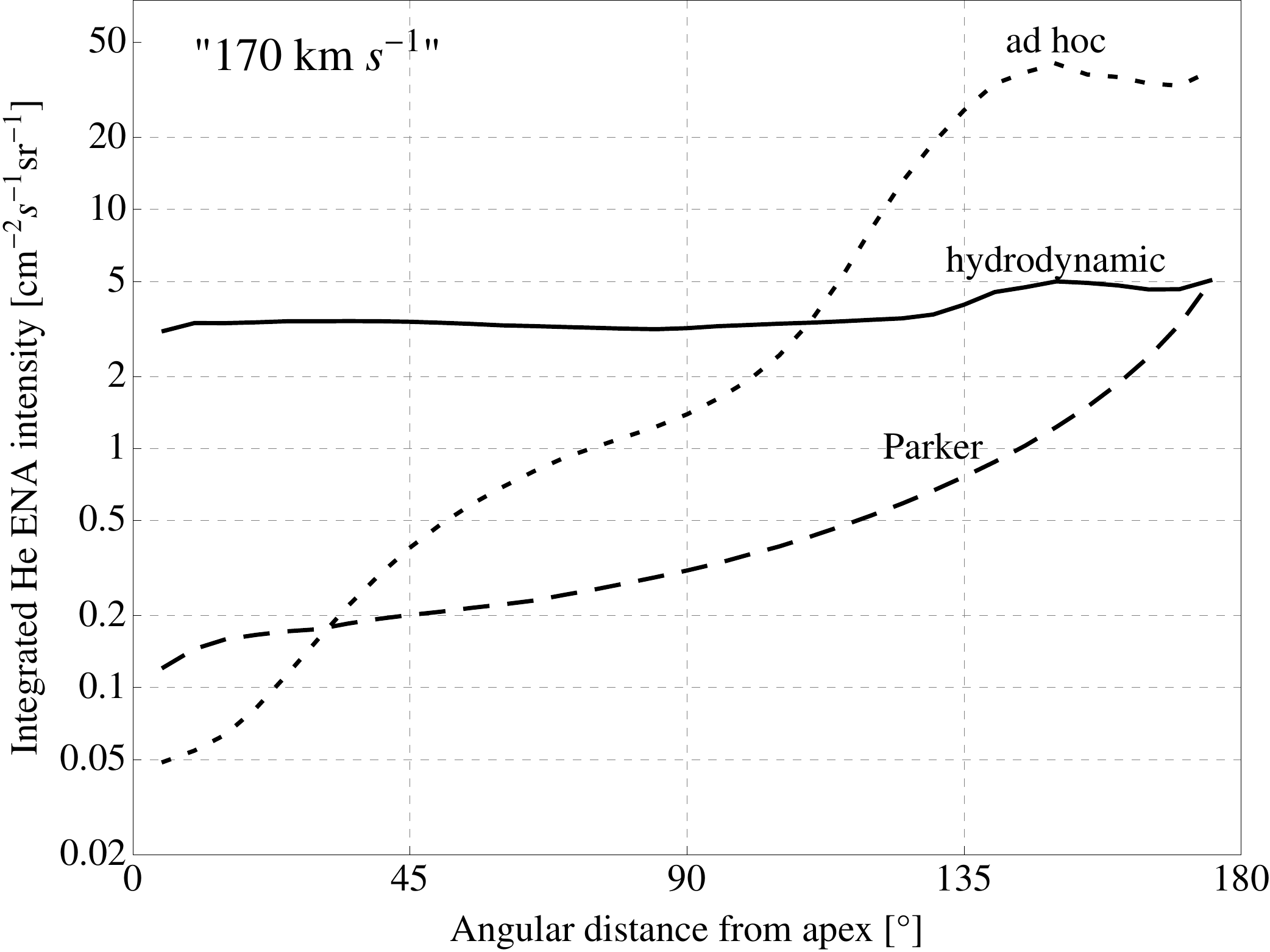}
   \includegraphics[width=0.48\textwidth]{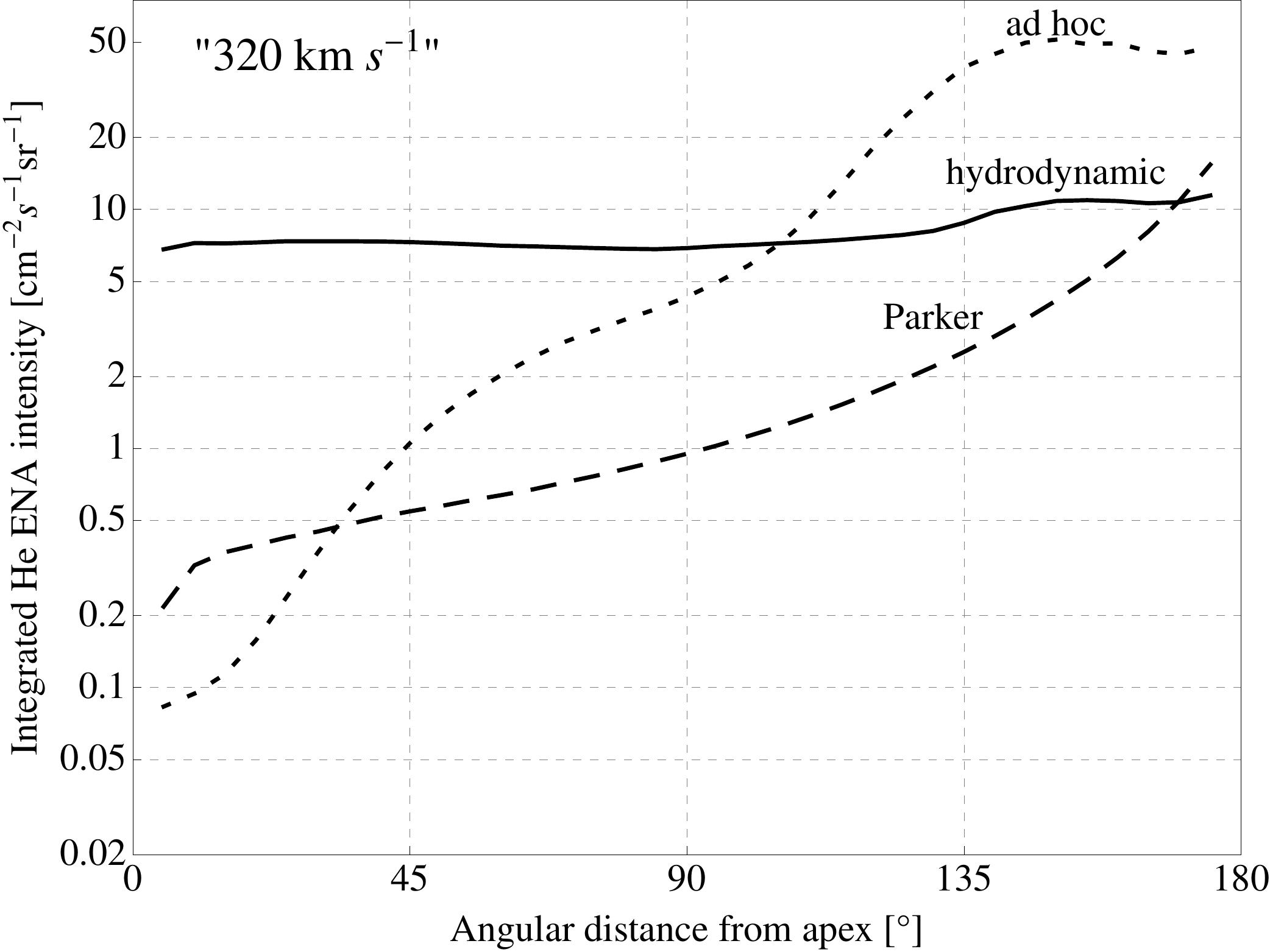}
   \caption{Total He ENA intensities (integrated over energy range 0.2 -- 50 keV and with losses included) as a function of angle $\theta$ from the apex for the hydrodynamic, Parker, and ad hoc models, and two variants of $\alpha$-particle injection at the TS: ``170~km~s$^{-1}$'' -- left panel and ``320~km~s$^{-1}$'' -- right panel (Sect.~\ref{aparticles}). He$^+$ injection is as described in Sect.~\ref{aparticles}.}
   \label{figintegrated}
\end{figure*}

The dependence of total He ENA intensity (integrated over energy range 0.2 -- 50~keV and with losses included) on angle $\theta$ from the apex is shown in Fig.~\ref{figintegrated} for the hydrodynamic, Parker, and ad hoc models for two assumed variants of $\alpha$-population injected at the TS (``170~km~s$^{-1}$'' and ``320~km~s$^{-1}$'') and He$^+$ injection as in Sect.~\ref{aparticles}. 

Results presented in Fig.~\ref{figintegrated} indicate that expected energy-integrated intensities of He ENA seem to depend more on the background model than to be sensitive to the energy of He ions injected to the inner heliosheath plasma at the TS. However, in all three models the intensity from the tail is greatest and ranges from a few ENA (cm$^{-2}$s$^{-1}$sr$^{-1}$) for the hydrodynamic and Parker models to a few tens of ENA (cm$^{-2}$s$^{-1}$sr$^{-1}$) for the ad hoc model. The fluxes from upwind are smaller than those from the tail, but the difference is marginal for the hydrodynamic model. The decrease is only by a factor $\sim$1.8. This is an effect of the He ion decharging relatively close to the Sun in the hydrodynamic model compared to other models. On the other hand, the tail/upwind contrast is significant for the Parker and ad hoc models (ratios of $\sim$60 and $\sim$800, respectively). 
Detection from the tail should therefore be the easiest, and the contrast tail/upwind may tell something about background plasma conditions.

\begin{figure*}
\centering
   \includegraphics[width=0.48\textwidth]{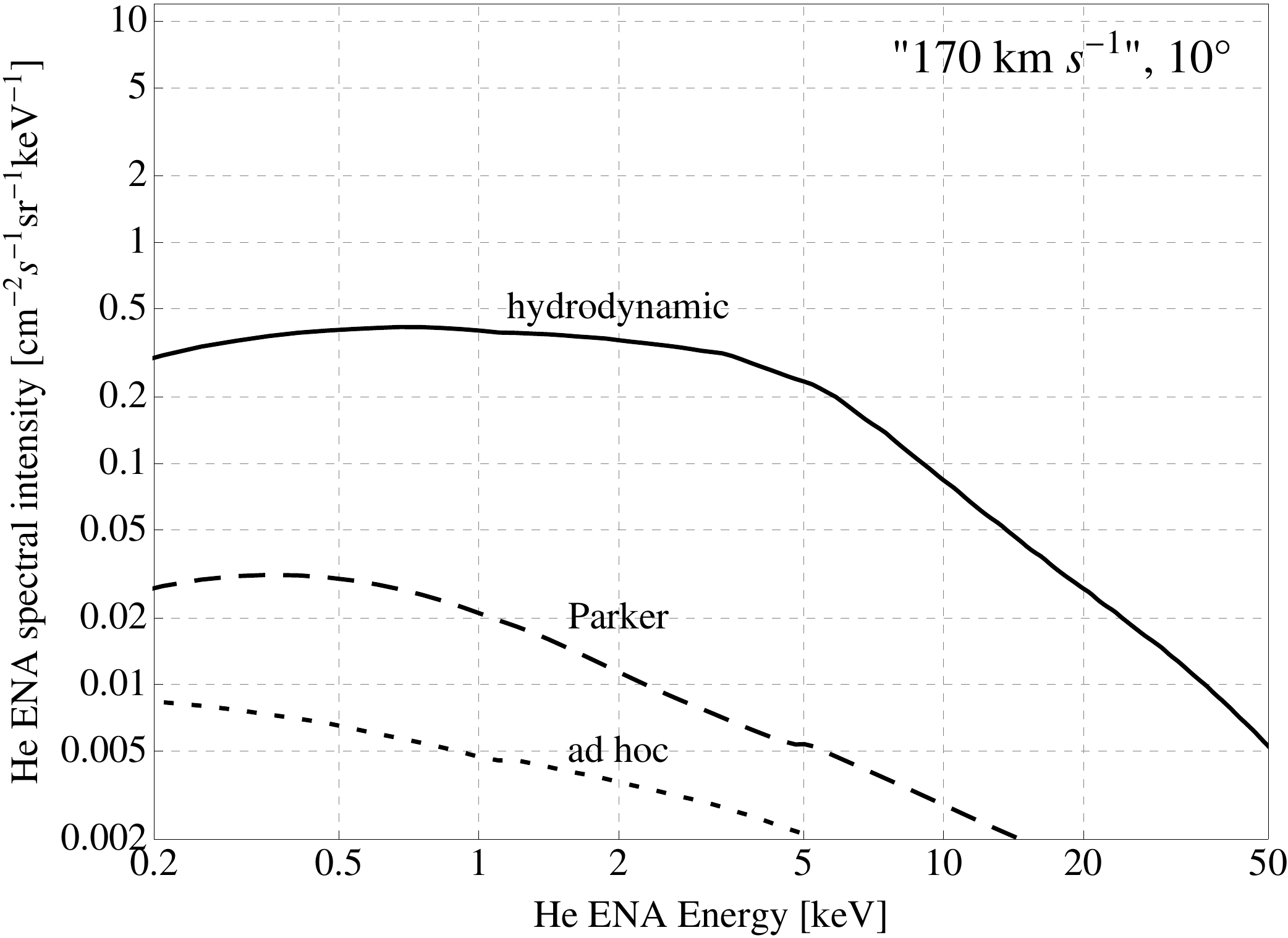}
   \includegraphics[width=0.48\textwidth]{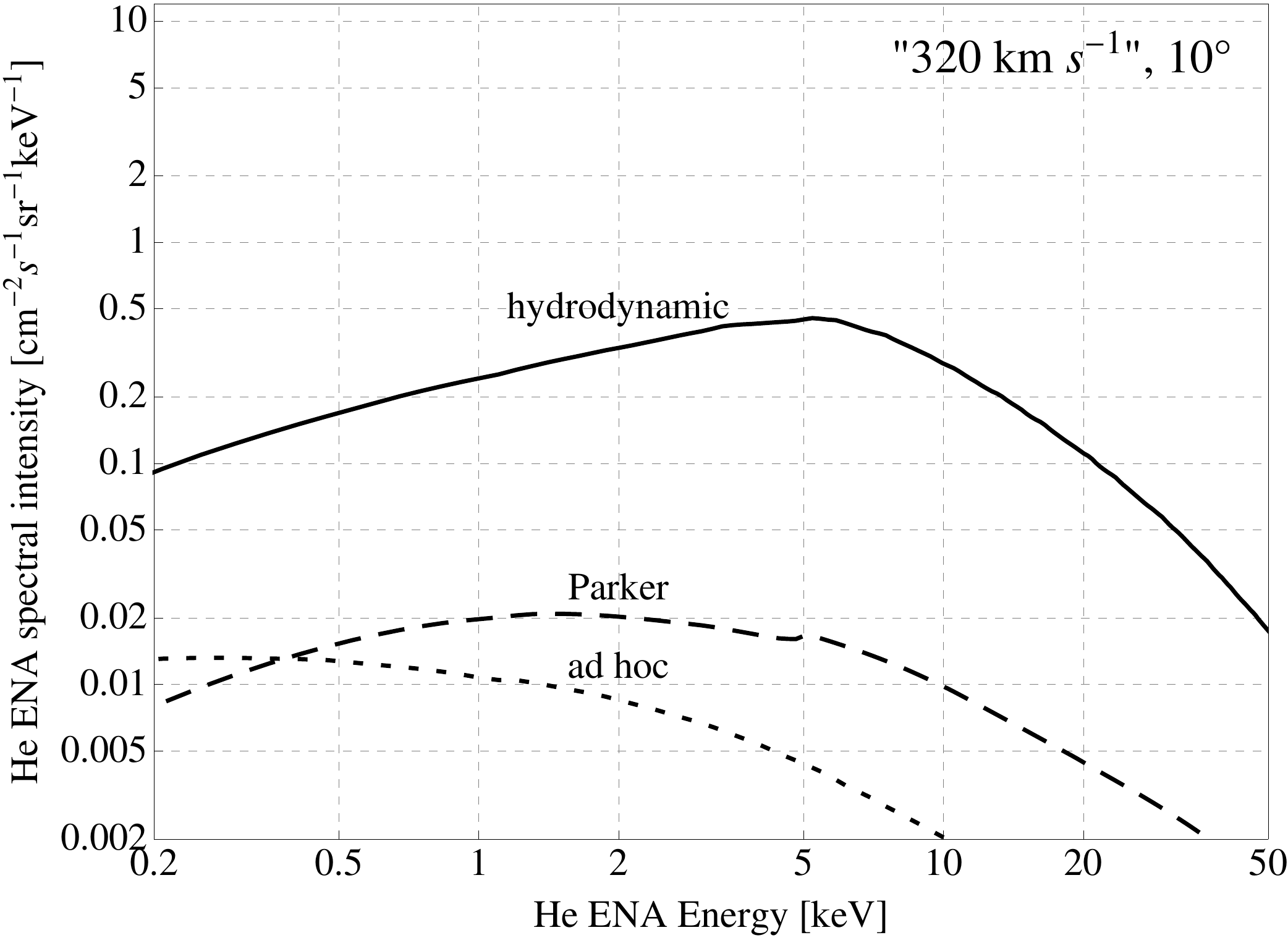}
   
   \includegraphics[width=0.48\textwidth]{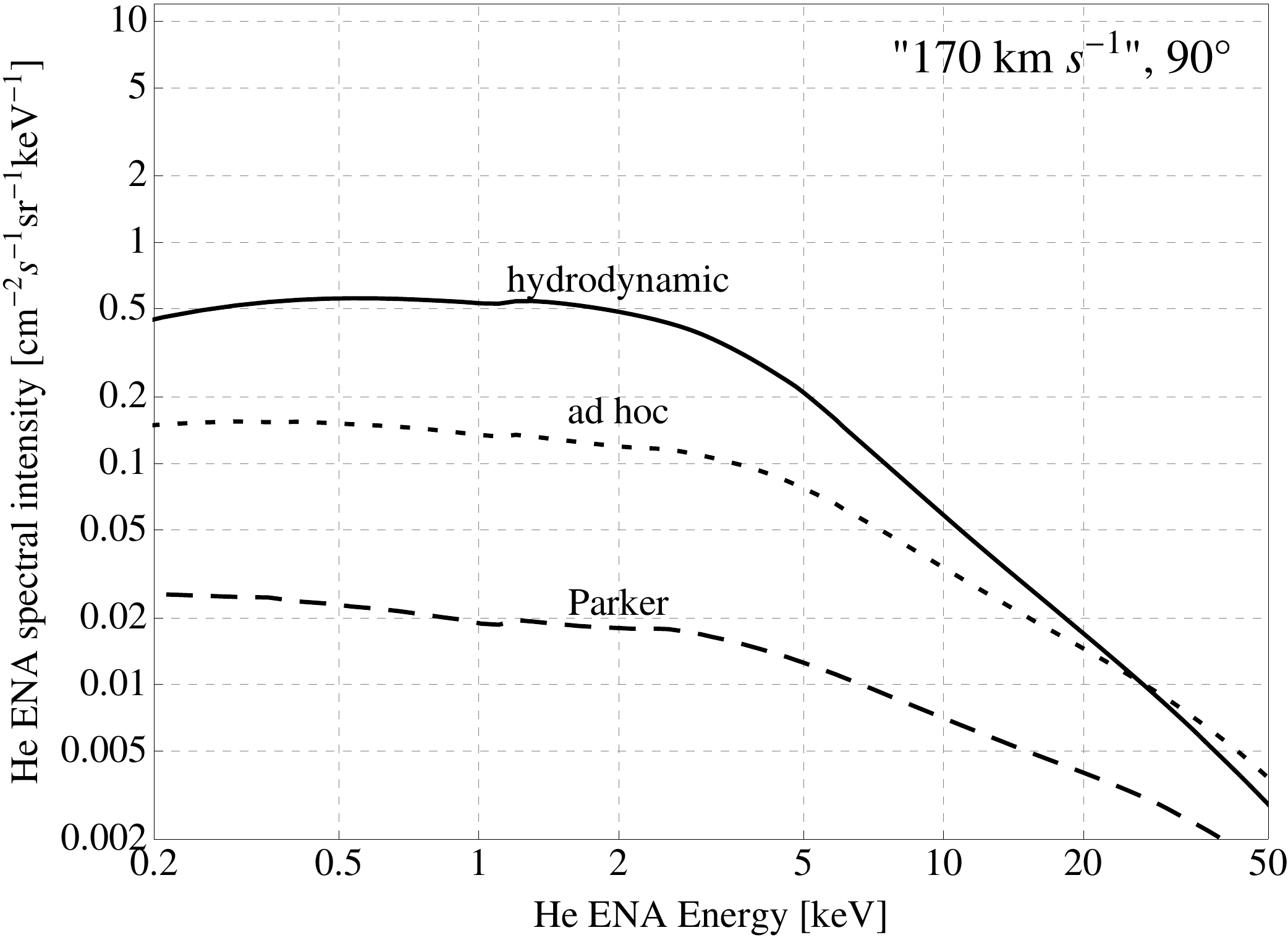}
   \includegraphics[width=0.48\textwidth]{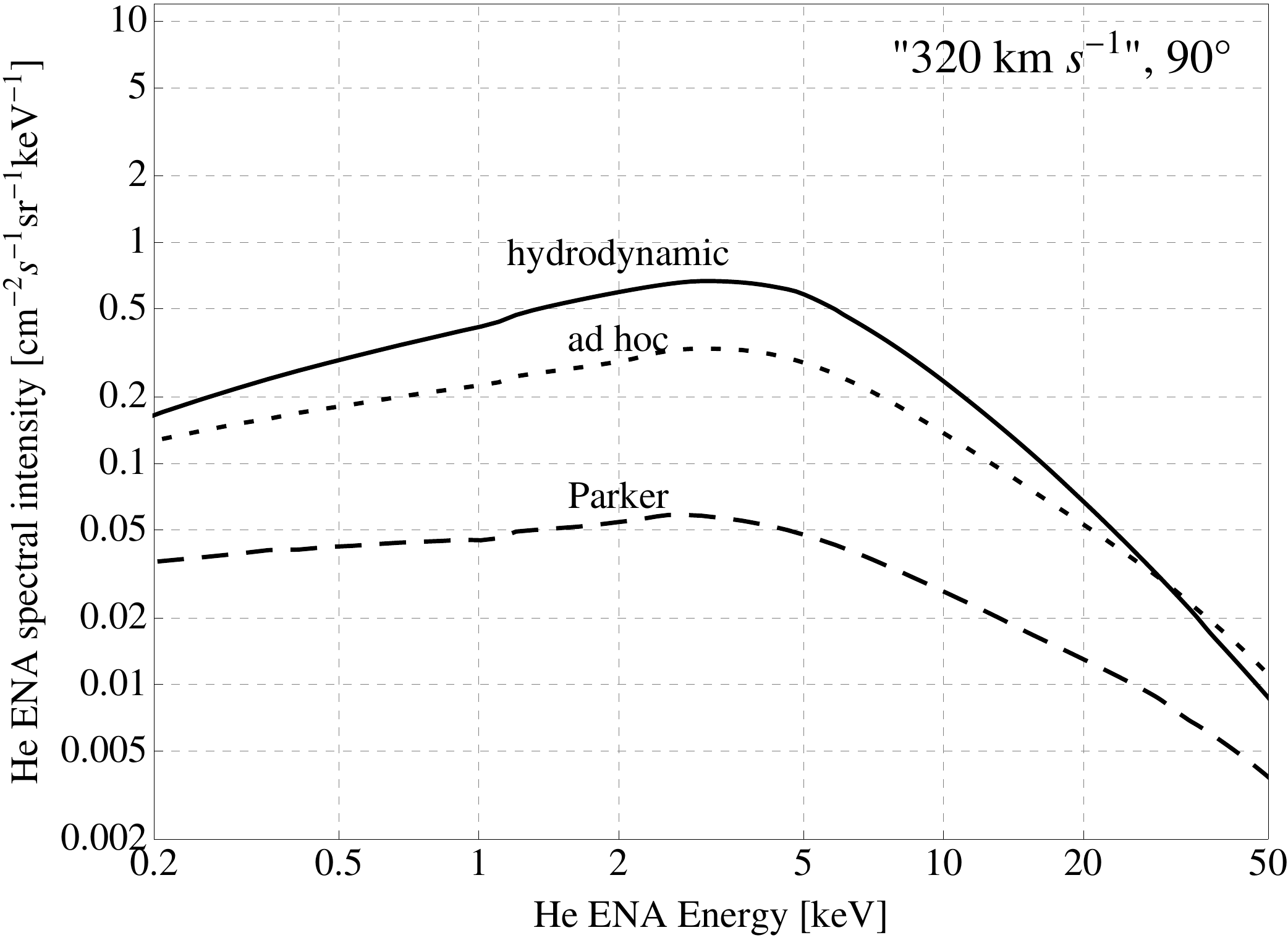}
   
   \includegraphics[width=0.48\textwidth]{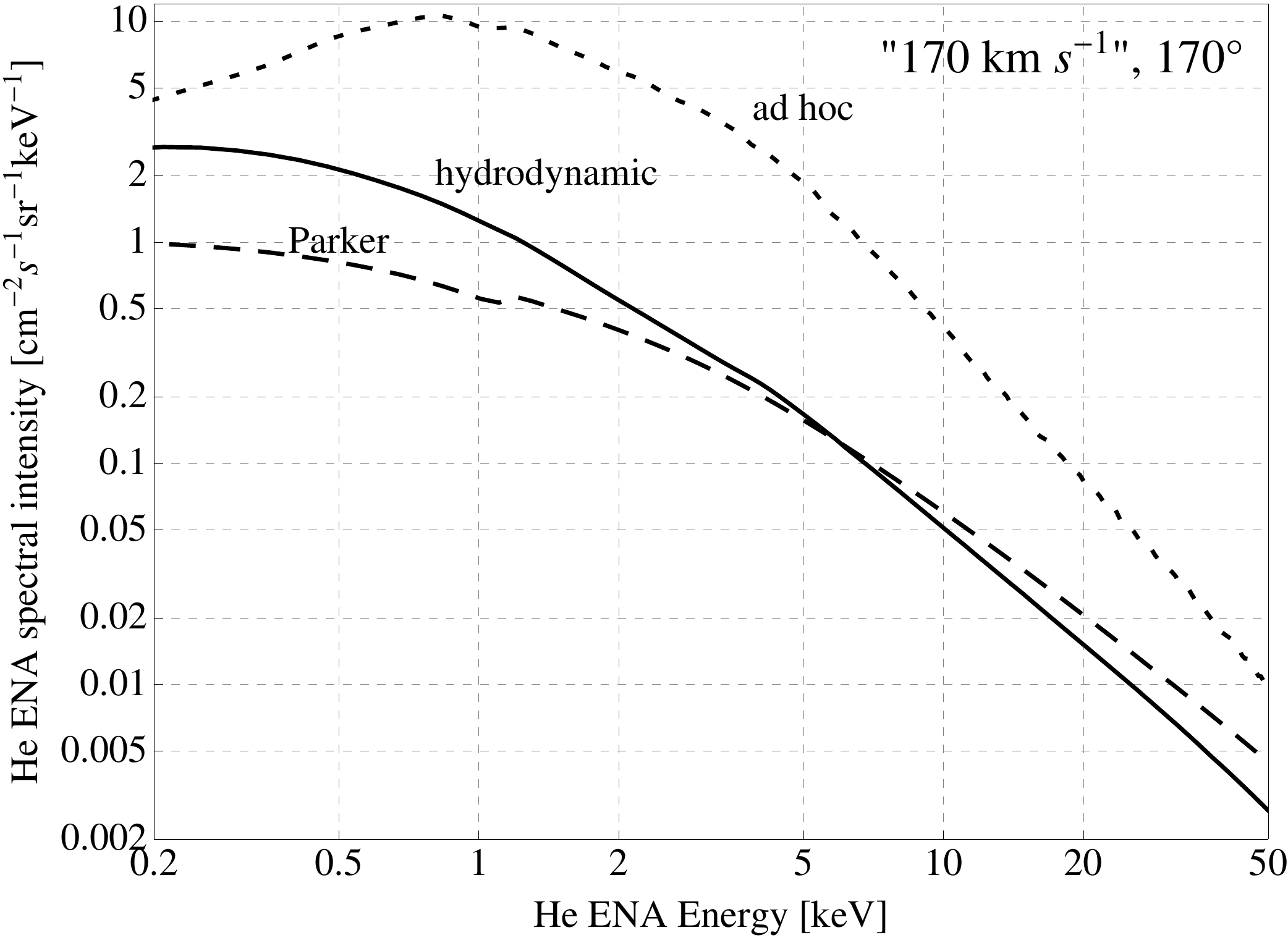}
   \includegraphics[width=0.48\textwidth]{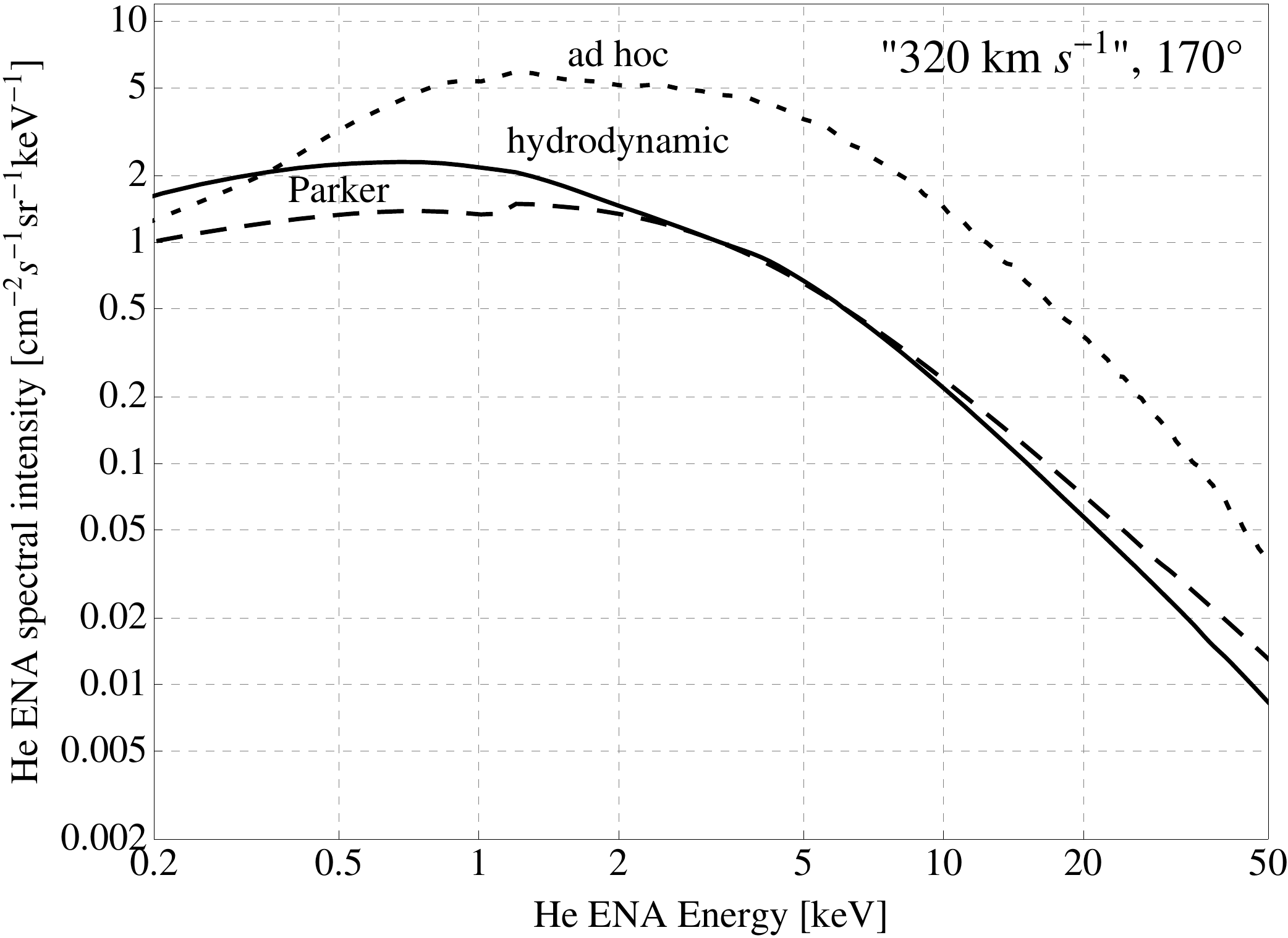}
   \caption{Expected energy spectra of He ENA (cm$^{-2}$s$^{-1}$sr$^{-1}$keV$^{-1}$) from the heliosheath. Left/right column -- TS injection variants ``170~km~s$^{-1}$''/``320~km~s$^{-1}$''. Upper/middle/lower row of panels -– LOS angle $\theta$ from apex 10$\degr$/90$\degr$/170$\degr$. For hydrodynamic/Parker/ad hoc models, the result is denoted by a solid/dashed/dotted line.}
   \label{figspectra}
\end{figure*}

Details on the dependence of He ENA energy spectra on model and angle $\theta$ are given in Fig.~\ref{figspectra} in which rows from top to bottom correspond to upwind ($\theta=10\degr$), crosswind ($\theta=90\degr$), and tail ($\theta=170\degr$) directions, for two variants of injection, ``170~km~s$^{-1}$'' (left panels) and ``320~km~s$^{-1}$'' (right panels). In all cases the spectra are broad in energy with maxima at $\lse$1~keV for ``170~km~s$^{-1}$'' and between $\sim$0.5 to $\sim$5~keV for ``320~km~s$^{-1}$''. Most energetic tail spectra are obtained in the ad hoc model. On the whole, the spectral region $\sim$0.5 -- $\sim$5~keV seems to be the most promising for detecting He ENA.

\section{Final discussion and conclusions}
\label{finaldiscussion}

Calculating the heliosheath production of ENAs out of heavy solar ions that capture electrons from neutral interstellar atoms is basically a straightforward exercise, if ion densities are estimated, the flux of interstellar neutrals is measured, and particle collision velocities can be calculated based on simple assumptions. This would apply if the state of background plasma were determined by hydrodynamics and if heavy ions (treated as test particles) were heavy enough to be weakly coupled to the background and therefore were able to preserve energies they had at the TS. Such an approach has been used in the past \citep{grzedzielski_etal:10a} to follow de-ionization of multicharged coronal ions of C, O, N, Mg, Si, and S carried by the solar wind. 

In the present paper that deals with light ions such as $\alpha$-particles and He$^+$, we have been faced with the situation that, on one hand, the interaction with the background plasma can be more important and, on the other, the evidently non-single fluid state of the background (as established by Voyager) affects various binary interactions to degrees that have not been assessed reliably. This can make the outcome very sensitive to details of heliosheath modeling. 

In view of the present lack of understanding what goes on in heliosheath plasmas, we tried to explore the range of expected He ENA fluxes for some simple models covering a wide gamut of possibilities. If a feature has emerged in all models, one may presume it has some relevance. 

As explained in Sect.~\ref{heliosheathbackground}, we employed three simple descriptions of heliosheath plasma background: hydrodynamic model by \citet{izmodenov_alexashov:03a}; classical Parker model; ad hoc model attempting to partially render the V1 \& V2 measured heliosheath data. These models cover a fairly wide range of possibilities. For all three models we followed the evolution of $\alpha$-particles and He$^+$ ions (isotropic) momentum distribution functions in phase space (Sects.~\ref{evolutionof}, \ref{binaryinteractions}, \ref{adiabaticheating}), calculated heliosheath density distributions and energy spectra (Sect.~\ref{aparticles}), and estimated He ENA energy fluxes expected at Earth (Sect.~\ref{expectedHe}). The last estimates are essentially the observables given in Table~\ref{tabsummary}. The data are organized following assumed peak injection energy of He-ions at the TS (cases ``170~km~s$^{-1}$'' and ``320~km~s$^{-1}$'', cf. Sect.~\ref{aparticles}).

\begin{table}
\caption{Expected energy-integrated He ENA intensities}
\label{tabsummary}   
\begin{center}                        
\begin{tabular}{l c c c}       
\hline\hline                 
Model 					& hydrodynamic 	& Parker 	& ad hoc 		\\    
					& (tail -- apex)& (tail -- apex)& (tail -- apex)	\\
\hline 
Case ``170 km~s$^{-1}$''		& 		& 		&			\\
intensity 				& 5 -- 3	& 5 -- 0.1	& 38 -- 0.05		\\
(cm$^{-2}$s$^{-1}$sr$^{-1}$) 		& 		& 		& 			\\
max. in spectrum			& 0.3 -- 0.7	& 0.2 -- 0.4	& 1 -- 0.2		\\
at  energy (keV)			& 		& 		&			\\
flux (cm$^{-2}$s$^{-1}$) 		& 25 -- 21	& 5 -- 1.5	& 95 -- 5 		\\
\\
Case ``320 km~s$^{-1}$''		& 		& 		&			\\
intensity				& 11 -- 7	& 15 -- 0.2	& 47 -- 0.1		\\
(cm$^{-2}$s$^{-1}$sr$^{-1}$) 		& 		& 		& 			\\
max. in spectrum 			& 0.7 -- 5	& 1.3 -- 1.5	& 3 -- 0.3		\\ 
at  energy (keV) 			& 		& 		&			\\
flux (cm$^{-2}$s$^{-1}$) 		& 54 -- 46	& 16 -- 4	& 150 -- 14 		\\
\hline                                   
\end{tabular}
\end{center}
\end{table}

IBEX is the first mission dedicated to the astronomy of energetic neutral atoms \citep{mccomas:09a}. It features two single-pixel, neutral atom cameras: IBEX-Hi \citep{funsten_etal:09a} and IBEX-Lo \citep{fuselier_etal:09a}. Basically only IBEX-Lo has the capability of discerning species of the incoming atoms, including (indirectly) neutral He \citep{mobius_etal:09a}, and IBEX-Hi was designed to observe neutral H. \citet{allegrini_etal:08a} suggest an ingeneous technique by which some other species, including He, might also be registered owing to a special treatment of data from the IBEX-Hi anti-coincidence system.

Results of this paper suggest that detecting maximum predicted fluxes might be within the reach of IBEX instrumentation. 
The full possibilities of He ENAs measurements have not as yet been published, but the possibility of detection depends critically on the ratio of He ENA to H ENA fluxes \citep{allegrini_etal:08a}. Based on published detection efficiencies \citep[Fig.~2]{allegrini_etal:08a} and H ENA fluxes given by \citet{mccomas:10a} we preliminarily estimate that He ENA intensities predicted by our modeling could be measured  from the heliospheric tail direction alone. For the ad hoc model predictions, the measurements should be possible in the IBEX-Hi energy channels 1.1, 2.7, and 4.3~keV, both for the ``170~km~s$^{-1}$'' and ``320~km~s$^{-1}$'' variants. For the hydrodynamic model predictions, the measurement might be marginally feasible at 2.7~keV for the ``320~km~s$^{-1}$'' variant.
In all considered models, maximum He ENA flux should come from the tail, though the effect may be only marginal for the hydrodynamic model (cf. fluxes given in Table~\ref{tabsummary}). If an angular tail-upwind decrease in the ENA signal could be measured, this might help to better pinpoint the essential physics of interactions. Maximum signal from the tail should also be related to maximum tail extension, which could help determine the direction of heliosphere wake, independently of the IBEX hydrogen data as in \citet{mccomas_etal:12a}, with obvious inferences concerning the interstellar medium state. 

Since the expected He ENA intensity at 1~AU from the Sun strongly depends on the conditions in the inner heliosheath, He ENA detection, if accomplished, may become an important asset in diagnosing the physical state of the innner heliosheath plasma with embedded He ions, as a complement to diagnosing via H ENA currently performed by IBEX. 

\bibliographystyle{aa} 
\bibliography{expected_fluxes_He_ENA}

\end{document}